\documentclass[pra,twocolumn,10pt,aps,showpacs,floatfix,amsmath,amsfonts,letterpaper,longbibliography]{revtex4-1}

\usepackage{stix} % Use Times fonts
\usepackage{graphicx}
\usepackage{amsmath}
\usepackage{url}
\usepackage{array}
\usepackage{multirow}
\usepackage{bm}

\newcolumntype{C}{>{\centering\arraybackslash}p{2.5cm}}
\newcolumntype{P}[1]{>{\centering\arraybackslash}p{#1}}

\makeatletter
\def\beq{\@ifstar{\@ifnextchar[{\@beqslabel}{\@beqsnolabel}}
{\@ifnextchar[{\@beqlabel}{\@beqnolabel}}}
\def\@beqlabel[#1]{\begin{equation}\label{#1}}
\def\@beqnolabel{\begin{equation}}
\def\@beqslabel[#1]{\begin{equation*}\label{#1}}
\def\@beqsnolabel{\begin{equation*}}
\def\eeq{\@ifstar{\end{equation*}}{\end{equation}}}
\makeatother

\newcommand{\refcite}[1]{Ref.~\cite{#1}}
\newcommand{\refcites}[1]{Refs.~\cite{#1}}

\newcommand{\refeq}[1]{Eq.~(\ref{#1})}
\newcommand{\refeqs}[2]{Eqs.~(\ref{#1})\nobreakdash--(\ref{#2})}
\newcommand{\refeqand}[2]{Eqs.~(\ref{#1}) and (\ref{#2})}

\newcommand{\reffig}[1]{Fig.~\ref{#1}}

\newcommand{\refsec}[1]{Sec.~\ref{#1}}
\newcommand{\refapp}[1]{Appendix~\ref{#1}}

\newcommand{\reftab}[1]{Table~\ref{#1}}

\newcommand{\etal}{\textit{et al.}}

\newcommand{\ee}{e}

\newcommand{\del}{\bm{\nabla}}

\newcommand{\punc}[1]{\,{\text{#1}}}
\newcommand{\sub}[1]{_{\text{#1}}}

\newcommand{\spr}[1]{^{(#1)}}

\newcommand{\rv}{\bm{r}}
\newcommand{\Rv}{\bm{R}}
\newcommand{\deltav}{\bm{\delta}}

\newcommand{\Mv}{\bm{M}}
\newcommand{\zerov}{\bm{0}}
\newcommand{\Phiv}{\bm{\Phi}}
\newcommand{\Bv}{\bm{B}}

\newcommand{\ham}{\mathcal{H}}

\usepackage[usenames,dvipsnames]{color}

\begin{document}

\title{Interacting double dimer model on the square lattice}

\author{Neil Wilkins}
\affiliation{School of Physics and Astronomy, The University of Nottingham, Nottingham, NG7 2RD, United Kingdom}

\author{Stephen Powell}
\affiliation{School of Physics and Astronomy, The University of Nottingham, Nottingham, NG7 2RD, United Kingdom}

\begin{abstract}
We study phases and transitions of the square-lattice double dimer model, consisting of two coupled replicas of the classical dimer model. As on the cubic lattice, we find a thermal phase transition from the Coulomb phase, a disordered but correlated dimer liquid, to a phase where fluctuations of the two replicas are closely synchronized with one another. Surprisingly, and in contrast to the cubic case, the phase boundary includes the noninteracting point, as we establish using a symmetry-based analysis of an effective height theory, indicating that infinitesimal coupling is sufficient to synchronize the double dimer model. In addition, we observe a novel antisynchronized phase when the coupling between replicas is repulsive, and use Monte Carlo simulations to establish the full phase diagram, including (anti)synchronized columnar and staggered phases, with interactions between parallel dimers in each replica.
\end{abstract}

\maketitle

\section{Introduction}
\label{introduction}

Symmetry-breaking phase transitions can be described using an `order parameter,' a local observable that transforms nontrivially under the symmetries, and which is strictly zero in the disordered phase but nonzero in the ordered phase \cite{Landau1980}. This precise distinction occurs only in the thermodynamic limit, since for a finite system thermal or quantum fluctuations necessarily restore the symmetry.

Among phase transitions to which this description does not apply, a prominent example is the Berezinskii--Kosterlitz--Thouless (BKT) transition in the two-dimensional (2D) XY model \cite{Berezinskii1971,Kosterlitz1973,Kosterlitz2016}. Across the BKT transition, no symmetry is broken, as required for a 2D system at nonzero temperature by the Mermin--Wagner--Hohenberg theorem \cite{Mermin1966,Hohenberg1967}, and there is hence no local order parameter. One can instead understand the BKT transition as an example of a `topological' phase transition \cite{Faulkner2015}, where the phases are distinguished by their topological properties. An appropriate criterion is the response to a twist applied across the boundaries of the system: the  associated energy cost, referred to as the helicity modulus (or phase stiffness), decreases exponentially with system size above the transition, but is nonzero in the thermodynamic limit below it \cite{Fisher1973,Nelson1977}.

\begin{figure}
\begin{center}
\includegraphics[width=0.65\columnwidth]{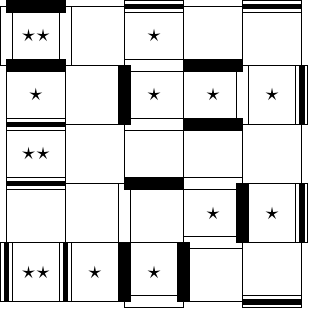}
\caption{An example configuration of the double dimer model on the square lattice, in which two replicas of the close-packed dimer model (shown in black and white) are defined on the same lattice. According to \refeq{eq:configurationenergy}, parallel pairs of nearest-neighbor dimers within each replica (marked with a star, \(\star\)) contribute $+J$ to the energy, and overlapping dimers contribute $+K$ to the energy. Hence, the energy of this configuration is $E = 15J + 9K$.}
\label{fig:model}
\end{center}
\end{figure}

In this work, we investigate a superficially distinct type of topological order in the square-lattice double dimer model, illustrated in \reffig{fig:model}, consisting of two coupled replicas of the standard dimer model. The dimer model  \cite{Kasteleyn1961,Temperley1961} is a paradigmatic example of a strongly correlated statistical system, in which the elementary degrees of freedom are dimers that occupy pairs of adjacent lattice sites subject to the constraint of close packing, i.e., that every site is occupied by exactly one dimer. This constraint, which can be interpreted as a Gauss law for an effective `magnetic field' \cite{Henley2010}, has profound consequences for the properties of the system.

In particular, the local Gauss law implies that the effective flux (defined by the lattice analogue of the standard magnetic flux) through any closed surface vanishes, while the flux across a surface spanning the system is conserved by any local rearrangement of dimers. The configuration space therefore splits into `topological sectors' corresponding to distinct values of the flux through each spanning surface, with a global rearrangement of dimers required to change sector.

This allows for the possibility of phases distinguished by the flux variance, which can be either suppressed exponentially or nonzero in the thermodynamic limit. The standard `single' dimer model on the square lattice is indeed known to exhibit a transition between phases showing these two behaviors, and at which spontaneous breaking of lattice symmetries occurs simultaneously \cite{Alet2005,Alet2006b}. Such transitions, even when accompanied by symmetry breaking, do not fit within the standard Landau paradigm \cite{Landau1980}, and the transition in the dimer model is in fact known to belong to the BKT universality class, with the flux variance playing the role of the helicity modulus \cite{Alet2005,Alet2006b}.

Our model contains interactions, of either sign, both between parallel dimers within each replica, \(J\), as studied previously in the single dimer model \cite{Alet2005,Alet2006b}, and between dimers that coincide (or `overlap') in the two replicas, \(K\). We show that it exhibits a particularly rich phase structure, including a number of phases with distinct types of order, both topological and symmetry-breaking.

As for the same model on the cubic lattice \cite{Wilkins2019}, we find a transition without any symmetry breaking, between a standard `Coulomb' phase \cite{Henley2010} and a `synchronized' phase, where both replicas remain disordered but their relative fluctuations are suppressed. More precisely, the difference between the fluxes in the two replicas has variance that becomes suppressed exponentially with system size, while each flux separately does not.

In contrast to the three-dimensional (3D) case, however, we find that this transition occurs for infinitesimal coupling \(K = 0^-\) between replicas, reflecting the critical nature of the noninteracting double dimer model \cite{Kenyon2014}. We also find a novel `antisynchronized' phase, where the overlap between replicas is minimized, which meets the Coulomb and synchronized phases at the zero-interaction point. Both of these features, along with the other transitions in the phase diagram, can be understood in terms of effective field theories based on `height models' \cite{Blote1982,Zeng1997}, which we derive based on symmetry.

For sufficiently strong repulsive (i.e., \(J > 0\)) interactions within each replica, we confirm the presence in the single dimer model of a transition into a staggered phase, as noted in previous works \cite{Alet2005,Castelnovo2007,Otsuka2009}, and determine the critical coupling at which it occurs. We also demonstrate the existence of phases in the double dimer model that are simultaneously staggered and (anti)synchronized.

Previous work on the double dimer model on the square lattice has addressed the noninteracting case \cite{Kenyon2014}, as well as models that correspond to the limit \(K \rightarrow +\infty\) \cite{Raghavan1997} (see \refsec{infinitecoupling}) and that include nonlocal interactions \cite{Damle2012}, motivated by a mapping from the quantum dimer model. Other related work has demonstrated the possibility of phase locking transitions in 2D superfluids \cite{Mathey2007} and the XY model \cite{Bighin2019}.

\subsection*{Outline}

In \refsec{model} we define the interacting double dimer model and present its phase diagram, which is calculated using the methods detailed in the subsequent sections: In \refsec{fieldtheoriesandcriticalproperties}, we use symmetries to write down height field theories that describe the various phases and transitions, before presenting, in \refsec{numericalresults}, the Monte Carlo (MC) results that underlie our phase diagram and establish the critical properties. We conclude in \refsec{conclusions}.

\section{Model}
\label{model}

We consider a classical statistical model of dimers on two replicas of an \(L\times L\) square lattice with periodic boundaries. To each link \(l\) of each replica \(\alpha \in \{1,2\}\), we assign a dimer occupation number \(d_l\spr{\alpha}\) which takes values \(0\) or \(1\). The close-packing constraint applies separately for each replica and requires that
\beq
\sum_{l \in \rv} d_l\spr{\alpha} = 1\punc,
\eeq
at each site \(\rv\), where the sum is over links \(l\) connected to \(\rv\).

To each configuration, we assign an energy
\beq[eq:configurationenergy]
E = J\left[N_{\parallel}\spr{1}+N_{\parallel}\spr{2}\right]+K\sum_{l} d_{l}\spr{1} d_{l}\spr{2}\punc,
\eeq
where \(J\) and \(K\) are, respectively, interaction strengths between parallel dimers within each replica and between overlapping dimers in the two replicas (see \reffig{fig:model}), and \(N_\parallel\spr{\alpha}\) counts the number of parallel pairs of nearest-neighbor dimers in replica \(\alpha\).   The partition function is given by \(Z = \sum \ee^{-E/T}\), where the sum is over all close-packed dimer configurations in both replicas. (We set \(k\sub{B} = 1\) throughout.)

\subsection{Magnetic field and height picture}
\label{heightpicture}

% Magnetic field
On the square lattice, it is useful to define a (fictitious) `magnetic field' \cite{Huse2003,Henley2010}
\beq[eq:magneticfield]
B_{\bm{r},\mu}\spr{\alpha} = \epsilon_{\rv}\left[d_{\rv,\mu}\spr{\alpha}-\frac{1}{q}\right]
\eeq
on the link joining sites \(\rv\) and \(\rv + \deltav_\mu\), where \(\deltav_\mu\) is a unit vector in direction \(\mu \in \{x,y\}\) (and the lattice spacing is set to \(1\)).
Here, $\epsilon_{\rv} = (-1)^{r_x + r_y} = \pm 1$ depending on the sublattice and $q=4$ is the coordination number. (A similar construction applies to other bipartite lattices such as the honeycomb lattice.) The close-packing constraint for the dimers is then equivalent to the condition that the `magnetic charge', given by the lattice divergence of \(B_{\rv,\mu}\spr{\alpha}\),
\beq[eq:defineQ]
Q\spr{\alpha}_{\rv} = \sum_\mu \left[ B_{\rv,\mu}\spr{\alpha} - B_{\rv-\deltav_\mu,\mu}\spr{\alpha} \right]\punc,
\eeq
is zero on every site.

% Height
In two dimensions, this divergence-free constraint is resolved by defining a scalar `height' \(z^{(\alpha)}\) on each plaquette, in terms of which
\begin{equation}
\label{eq:height}
B_{\bm{r},\mu}^{(\alpha)} = \epsilon_{\mu\nu} \Delta_{\nu} z^{(\alpha)} \punc ,
\end{equation}
where $\epsilon_{\mu\nu}$ is the two-dimensional Levi-Civita symbol and $\Delta_{\nu}$ denotes the lattice derivative \cite{Blote1982, Zeng1997}. (This is the two-dimensional analog of $\bm{B} = \bm{\nabla} \times \bm{A}$.)

\begin{figure}
\begin{center}
\includegraphics[width=0.99\columnwidth]{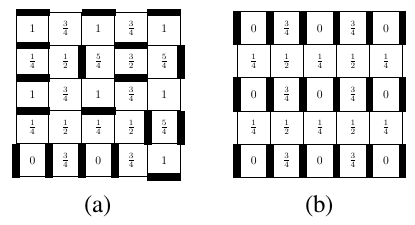}
\caption{Example height representations $z$ of (a) an arbitrary dimer configuration and (b) the columnar configuration with magnetization order parameter \(\Mv = \deltav_y\), which has average height $\langle z \rangle = \frac{3}{8}$.}
\label{fig:height}
\end{center}
\end{figure}

% Rules
Together, \refeqand{eq:magneticfield}{eq:height} define a one-to-one mapping between dimer configurations and their height representations, which is usually expressed as the following set of rules \cite{Alet2006b}: One first chooses a plaquette to be the zero of height. Then, moving anticlockwise around sites on sublattice A (B), the height increases (decreases) by $1 - 1/q$ when an occupied bond is crossed. If, instead, an empty bond is crossed, the height decreases (increases) by $1/q$. Example height representations are shown in \reffig{fig:height}.

% Flux
The flux \(\Phiv\spr\alpha\) for each replica \(\alpha\) can be defined by
\beq[eq:fluxdef]
\Phi_{\mu}\spr\alpha = \frac{1}{L}\sum_{\rv}B_{\rv,\mu}\spr\alpha = \frac{1}{L}\sum_{\rv}\epsilon_{\bm{r}}d_{\rv,\mu}\spr\alpha
\punc,
\eeq
which, because of the divergence-free constraint, is equivalent to the sum of the magnetic fields on links crossing a surface normal to \(\deltav_\mu\). The latter definition implies that \(\Phi_{\mu}\spr\alpha\) takes integer values and furthermore that it can be changed only by shifting dimers around a loop encircling the whole system \cite{Chalker2017}.

\subsection{Phase diagram}
\label{phasediagram}

\begin{figure}
\begin{center}
\includegraphics[width=\columnwidth]{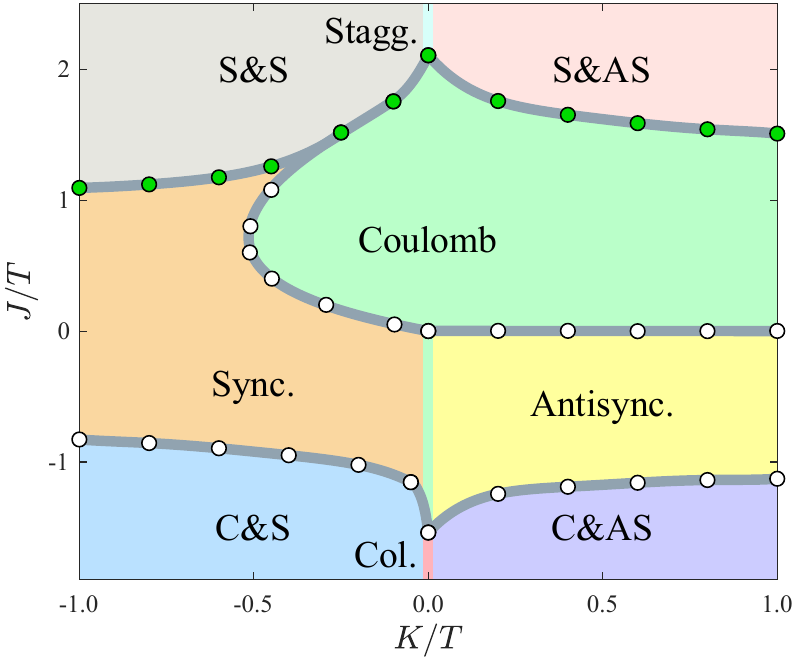}
\caption{Phase diagram for the double dimer model of \refeq{eq:configurationenergy} on the square lattice, in the $(J/T,K/T)$ plane. Dots show points where
the phase boundary has been determined using an RG analysis ($J=K=0$) or MC simulations (all other points), and thick grey lines are guides to the eye. The ordered phases are: columnar (`Col'), staggered (`Stagg.'), synchronized (`Sync.'), antisynchronized (`Antisync.'), columnar \& synchronized (`C\&S'), columnar \& antisynchronized (`C\&AS'), staggered \& synchronized (`S\&S') and staggered \& antisynchronized (`S\&AS'). White dots represent BKT transitions, while green dots represent apparently continuous transitions.}
\label{fig:phasediagram}
\end{center}
\end{figure}

Our phase diagram for the square-lattice double dimer model, \refeq{eq:configurationenergy}, is shown in \reffig{fig:phasediagram}. The fact that the Coulomb, synchronized and antisynchronized phases meet at $K=J=0$ (in contrast to the cubic-lattice case; see  Fig.~3 of \refcite{Wilkins2019}) is determined solely from an RG analysis in \refsec{doubledimermodel}. All other points are obtained numerically using a MC worm algorithm \cite{Sandvik2006, Wilkins2019}, as we describe in \refsec{numericalresults}. In the remainder of this section, we define the phases appearing in \reffig{fig:phasediagram}.

\subsubsection{Independent replicas}
\label{independentreplicas}

For \(K=0\) the two replicas are independent and behave as single dimer models with interactions that favor (\(J < 0\)) or disfavor (\(J > 0\)) parallel dimers. For \(J = 0\), this model exhibits a Coulomb phase \cite{Henley2010}, where no symmetries are broken and the connected dimer--dimer correlation function \(\langle d_l d_{l'} \rangle\sub{c}\) decreases algebraically with separation. This phase extends to small nonzero \(J\), but gives way to ordered phases for sufficiently large \(\lvert J \rvert/T\).

\begin{figure}[!t]
\begin{center}
\includegraphics[width=\columnwidth]{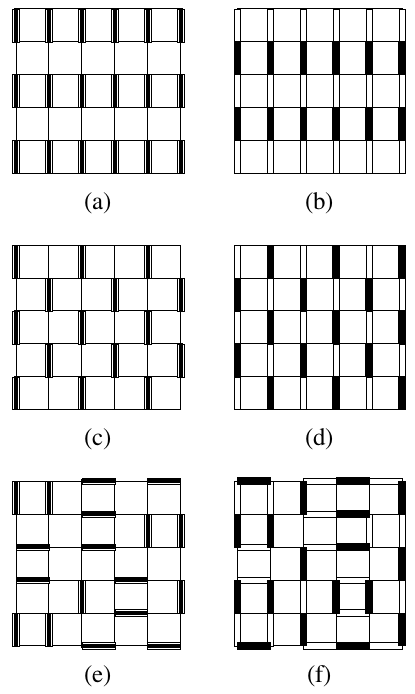}
\caption{Example ground states of the double dimer model of \refeq{eq:configurationenergy} on the square lattice. (a)--(b) Columnar configurations, with maximal number of parallel plaquettes $N_{\parallel}\spr{\alpha}$, which minimize the energy for $J<0$, $K=0$. For $J<0$, $K<0$ configuration (a), with maximal overlap, is a columnar \& synchronized ground state. Configuration (b) is a columnar \& antisynchronized ground state when $J<0$, $K>0$, because it has zero overlap between replicas. (c)--(d) Staggered configurations, with zero parallel plaquettes $N_{\parallel}\spr{\alpha}$, which minimize the energy for $J>0$, $K=0$. For $J>0$, $K<0$, configuration (c) is a staggered \& synchronized ground state, while for $J>0$, $K>0$, configuration (d) is a staggered \& antisynchronized ground state. (e) A fully synchronized configuration, which is a ground state for $J=0$, $K<0$. (f) A fully antisynchronized configuration which is a ground state for $J=0$, $K>0$.}
\label{fig:groundstates}
\end{center}
\end{figure}

For negative \(J\), there is a transition to a phase with columnar order, as illustrated in Figs.~\ref{fig:groundstates}(a) and (b), breaking translation and rotation symmetries \cite{Alet2005,Alet2006b}. An appropriate order parameter for this phase is the `magnetization'
\beq[eq:Mdef]
M_\mu = \frac{2}{L^2}\sum_{\rv} (-1)^{r_\mu} d_{\rv,\mu}\punc,
\eeq
which takes the values \(\Mv = \pm \deltav_\mu\) in the four columnar states that maximize \(N_\parallel\).

% LaTeX struggles to place this large figure, so use "!t" to allow it to break the placement rules.

Besides the symmetry-breaking order parameter, the two phases are also distinguished by the probability distribution \(P(\Phiv)\) for the flux \(\Phiv\). In the thermodynamic limit, the Coulomb phase has \(P(\bm{\Phi}) \propto \ee^{-\frac{\kappa}{2}\lvert\bm{\Phi}\rvert^{2}}\), where \(\kappa\) is a function of \(J/T\) (see \refapp{sdmfluxinthecoulombphase}). In the ordered phase, by contrast, \(P(\Phiv)\) is suppressed exponentially with system size for nonzero \(\Phiv\), since changing the flux requires shifting a row of dimers that spans the whole system, with energy cost proportional to \(L\). The mean square flux \(\langle\lvert\Phiv\rvert^2\rangle\) therefore changes its behavior across the transition, being independent of \(L\) in the Coulomb phase but vanishing in the thermodynamic limit in the columnar phase \cite{Alet2006b}.

We also make use of a third diagnostic of the transition, which is based on confinement of `monomers', empty sites in the otherwise close-packed configuration. Removing one dimer leaves a pair of monomers on adjacent sites, with unit charges \(Q_{\rv} = \pm 1\), which one can separate by locally rearranging the remaining dimers \cite{Henley2010}. We define the monomer distribution function \(G\sub{m}(\rv_+ - \rv_-) = Z\sub{m}(\rv_+,\rv_-)/Z\), where \(Z\sub{m}\) is the sum of Boltzmann weights of all configurations with a pair of monomers fixed at \(\rv_+\) and \(\rv_-\).

In the Coulomb phase, \(G\sub{m}\) decreases algebraically with separation (see \refapp{sdmgminthecoulombphase}), corresponding to a logarithmic effective potential \(U\sub{m}(\Rv) \equiv -\ln G\sub{m}(\Rv) \sim \ln \lvert\Rv\rvert\) \cite{Chalker2017}. In the columnar phase, separating the monomers disturbs the ordered configuration, causing a linear potential \(U\sub{m}(\Rv) \sim \lvert\Rv\rvert\), and so \(G\sub{m}(\Rv)\) decreases exponentially with \(\lvert\Rv\rvert\). The potential \(U\sub{m}\) therefore grows without limit in both phases; this is in contrast with the 3D case, where the potential is bounded in the Coulomb phase, and the monomers are said to be `deconfined' \cite{Henley2010}. The different asymptotic behaviors nonetheless allow the phases to be distinguished, and we refer to the 2D Coulomb phase as `quasideconfined' by analogy with quasi-long-range order in the low-temperature phase of the XY model \cite{Kosterlitz2016}.

For large positive \(J\), the system instead reduces the number of parallel dimers. The square lattice has a large but subextensive set of `staggered' configurations with the minimal value \(N_\parallel = 0\) and \(\Phiv\) of order \(L\) [see, for example, Figs.~\ref{fig:groundstates}(c) and (d)]. As a result, there is a transition at large positive \(J/T\) to a phase where the flux vector takes a nonzero expectation value of order \(L\) \cite{Alet2005,Castelnovo2007,Otsuka2009}. We treat this phase and the transition in detail in \refsec{staggeredorderingtransitions}.

The Coulomb, columnar, and staggered phases of the single dimer model are shown in \reffig{fig:phasediagram} on the vertical line \(K/T = 0\). Note that the Coulomb phase extends to \(J < 0\) along the line \(K/T = 0\).

\subsubsection{Coupled replicas}
\label{coupledreplicas}

For \(K \neq 0\), the two replicas are coupled, with overlapping dimers favored for \(K < 0\) and disfavored for \(K > 0\).

The columnar and staggered phases at large \(\lvert J\rvert /T\) have order parameters, \(\Mv\spr{\alpha}\) and \(\Phiv\spr{\alpha}\) respectively, in each replica \(\alpha\). In the thermodynamic limit, any nonzero coupling \(K\) fixes the relative values in the two replicas in order to maximize or minimize the overlap, as illustrated for the ground states in Figs.~\ref{fig:groundstates}(a)--(d). We refer to the resulting phases as columnar/staggered \& (anti)synchronized. In particular, order by disorder selects $\bm{M}\spr1=-\bm{M}\spr2$ in the columnar \& antisynchronized phase \cite{Wilkins2019}, while $\bm{\Phi}\spr1 = -\bm{\Phi}\spr2$ in the staggered \& antisynchronized phase (see \refsec{staggeredorderingtransitions}).

For smaller values of \(\lvert J\rvert /T\), phase transitions occur that do not involve symmetry breaking, but can be characterized through the flux distribution and monomer confinement.

The flux distribution in the double dimer model can be described by the \(2\times 2\) covariance matrix \(\langle \Phiv\spr{\alpha} \cdot \Phiv\spr{\alpha'}\rangle\), but symmetry under replica exchange means we need only consider \(\langle\lvert \Phiv\spr{\pm}\rvert^2\rangle\), where \(\Phiv\spr{\pm} = \Phiv\spr{1} \pm \Phiv\spr{2}\) are the total and relative fluxes corresponding to the fields \(B\spr\pm = B\spr1 \pm B\spr2\). Note that the double dimer model can be viewed as a model of directed loops in \(B\spr-\), formed from the overlap of the two replicas \cite{Wilkins2019}, and that \(\Phiv\spr{-}\) is the loop flux in this picture.

For \(K < 0\), one can postulate a phase where both replicas remain disordered and their flux variances \(\Phiv\spr\alpha\) remain nonzero, but where their fluctuations are strongly correlated so that the variance of the relative flux \(\langle\lvert\Phiv\spr-\rvert^2\rangle\) vanishes in the thermodynamic limit [see \reffig{fig:groundstates}(e)]. We have previously identified such a phase, which we call `synchronized', in the double dimer model on the cubic lattice \cite{Wilkins2019}, and we demonstrate in the following that it also occurs on the square lattice. For \(K > 0\), we similarly identify an `antisynchronized' phase, where fluctuations are correlated between the replicas in order to reduce the amount of overlap [see \reffig{fig:groundstates}(f)]. The relative flux \(\langle\lvert\Phiv\spr-\rvert^2\rangle\) also vanishes in the antisynchronized phase, as we argue in \refsec{doubledimermodel}.

The monomer-confinement criterion can also be applied in the double dimer model, where we define \(G\sub{m}\) using a pair of monomers of opposite charge in the same replica, say \(\alpha = 1\). Each monomer then has nonzero charge for \(B\spr1\) and hence for both \(B\spr-\) and \(B\spr+\). They are therefore confined, with \(G\sub{m}(\Rv)\) decreasing exponentially with \(\lvert\Rv\rvert\), in the (anti)synchronized phases, where fluctuations of \(\Phiv\spr-\) are suppressed.

To distinguish the columnar-ordered phases from the (anti)synchronized phases, one can instead insert pairs of monomers in both replicas simultaneously. Two monomers, one in each replica, on the same lattice site form a double charge for \(B\spr+\), but have zero net charge for \(B\spr-\). We therefore expect such objects to be confined only when the total flux variance is suppressed. Explicitly, we define the double monomer distribution function as \(G\sub{d}(\rv_+ - \rv_-) = Z\sub{d}(\rv_+,\rv_-)/Z\), where \(Z\sub{d}\) is the sum of Boltzmann weights of all configurations with a pair of monomers fixed at \(\rv_+\) and \(\rv_-\) in each replica. In the columnar-ordered phases, $G\sub{d}(\Rv)$ decreases exponentially with $\lvert \Rv \rvert$, whereas in both the (unsynchronized) Coulomb phase and the (anti)synchronized phases, \(G\sub{d}(\Rv)\) decreases only algebraically with \(\lvert\Rv\rvert\). [For the (anti)synchronized phases, we show this directly in \refapp{ddmgdinthe(anti)synchronizedphases} using an effective field theory.]

\subsubsection{Infinite coupling between replicas}
\label{infinitecoupling}

The point \(J=0\), \(K/T \rightarrow +\infty\) corresponds to the dimer loop model \cite{Raghavan1997}, which is equivalent to a fully-packed loop model with fugacity \(n=2\). The latter is known to be nonintegrable on the square lattice \cite{Jacobsen2004,Zinn-Justin2009} but solvable on the honeycomb lattice, where it is equivalent to a three-coloring model \cite{Baxter1970}.

In the opposite limit \(K/T\rightarrow-\infty\), the two replicas are perfectly aligned, and so act as a single dimer model with coupling \(2J\) between parallel dimers. The values of \(J/T\) at the columnar and staggered phase boundaries in this limit are therefore exactly half their values at \(K=0\). (For \(K/T\rightarrow+\infty\), the critical couplings lie in between these two values, because the dimer loop model has higher entropy than the single dimer model.)

\section{Field theories and critical properties}
\label{fieldtheoriesandcriticalproperties}

Using the height mapping, the long-wavelength properties of the Coulomb phase can be described in terms of a free field theory. In this section, we use symmetry to determine the perturbations to this action that are most relevant under the renormalization group (RG), and hence establish the critical properties at each transition.

\subsection{Single dimer model}
\label{singledimermodel}

% Coulomb phase, single dimer model
To construct a continuum theory we replace the effective magnetic field \(B_{\rv,\mu}\) and height $z$ by coarse-grained fields \(\Bv(\bm{r})\) and $h(\bm{r})$ obeying \(B_\mu(\bm{r}) = \epsilon_{\mu\nu}\partial_\nu h(\bm{r})\). For the single dimer model, the Coulomb phase has action
\beq
\label{eq:SDMCoulomb}
S\sub{SDM} = \int d^{2}\bm{r} \, \frac{\kappa}{2}\lvert \bm{B} \rvert^2 = \int d^{2}\bm{r} \, \frac{\kappa}{2}\lvert \del h \rvert^2 \punc,
\eeq
where $\kappa$ is the stiffness, plus irrelevant higher-order terms. In the non-interacting limit ($J = 0$) the stiffness is
\begin{equation}
\label{eq:exactstiffness}
\kappa_{\infty} = \pi \punc ,
\end{equation}
from comparison of observables, for example \refeqand{eq:fluxCoulomb}{eq:heightgm}, with exact results obtained using Pfaffian methods \cite{Boutillier2009,Fisher1963}.

% Columnar-ordering transition action
To study the columnar-ordering transition in the single dimer model, we include additional terms in \refeq{eq:SDMCoulomb}. We require that any action is local, and invariant under both $\pi/2$ rotations and translation of dimers; as discussed in \refcite{Alet2006b}, this imposes constraints on the form of allowed additional terms, which are summarized in \reftab{table:constraints}. For example, translation of dimers by one lattice constant in the $x$ direction maps the height field $h(\rv) \rightarrow -h(\rv - \deltav_{x}) - \frac{1}{4}$, so the action must be invariant under this change. The critical theory, which includes the most relevant term (in the RG sense) consistent with all requirements, is a sine-Gordon model:
\beq
\label{eq:SDMcol}
S\sub{SDM,col.} = S\sub{SDM} + \int d^{2}\bm{r}\,V\cos(8\pi h).
\eeq
Note that if the symmetry of the single dimer model is reduced \cite{Chen2009}, such as in the case of anisotropic interaction strengths, i.e., $J_{x} \neq J_{y}$, between parallel dimers \cite{Otsuka2009}, the form of the allowed cosine term is modified.

\renewcommand{\arraystretch}{1.2}
\begin{table*}
\begin{center}
\begin{tabular}{ | P{3.5cm} | P{3.5cm} | P{3.5cm} | P{3.5cm} | }
\hline
\multirow{2}{*}{Requirement} & \multicolumn{3}{|c|}{Constraint} \\ \cline{2-4}
& SDM & coupled DDM, $h\spr-$ & coupled DDM, $h\spr+$ \\ \hline
Locality & $S[h] =S[h + 1]$ & $S[h\spr-] =S[h\spr- + 1]$ & $S[h\spr+] =S[h\spr+ + 1]$ \\ \hline
$\pi/2$ rotation symmetry & $S[h] = S[-h]$ & $S[h\spr-] = S[-h\spr-]$ & $S[h\spr+] = S[-h\spr+]$ \\ \hline
Translation symmetry & $S[h] = S[-h - 1/4]$ & $S[h\spr-] = S[-h\spr-]$ & $S[h\spr+] = S[-h\spr+ - 1/2]$ \\
\hline
\end{tabular}
\caption{Requirements for the action of the single dimer model (SDM) and the double dimer model (DDM) with coupled replicas on the square lattice, and their corresponding constraints on allowed additional terms. (In the case of the DDM with independent replicas, SDM constraints apply separately to both $h\spr1$ and $h\spr2$.) Here, $\pi/2$ rotation symmetry refers to rotations about a plaquette center, and translation symmetry refers to translations by one lattice constant in the $x$ direction. For the SDM `$S$' means $S\sub{SDM,col.}$, which describes the columnar-ordering transition [see \refeq{eq:SDMcol}]. For the coupled DDM `$S$' means $S\sub{DDM,sync.}$, which describes the synchronization transition [see \refeq{eq:DDMsync}], or $S\sub{DDM,col.}$, which describes columnar ordering of coupled replicas [see \refeq{eq:DDMcol}]. The SDM constraints on $S[h]$ are discussed in detail in \refcite{Alet2006b}, and can be used to deduce the coupled DDM constraints on $S[h\spr\pm]$.}
\label{table:constraints}
\end{center}
\end{table*}
\renewcommand{\arraystretch}{1}

% Perturbative RG calculation
A standard perturbative RG calculation \cite{Alet2006b} applied to the general sine-Gordon action
\begin{equation}
S\sub{SG} = \int d^{2}\bm{r} \left[ \, \frac{\kappa}{2} \lvert \del h \rvert^2 + V\cos(2\pi p h) \right] \punc ,
\end{equation}
with $p$ an integer, leads to the following conclusions: There is a BKT phase transition at a critical value of the stiffness
\begin{equation}
\label{eq:rg}
\kappa\sub{c} = \frac{1}{2}\pi p^{2} \punc .
\end{equation}
When $\kappa < \kappa\sub{c}$ the cosine term is irrelevant, i.e., it renormalizes to zero in the long distance theory, which is thus a free Coulomb phase. When $\kappa > \kappa\sub{c}$ it is relevant and locks the height field to discrete values.

% Columnar-ordering transition discussion
In the case of the columnar-ordering transition where the action, \refeq{eq:SDMcol}, has $p=4$, we have
\begin{equation}
\label{eq:kappac}
\kappa\sub{c} = 8\pi \punc .
\end{equation}
In the columnar phase ($\kappa > \kappa\sub{c}$) the cosine term locks the height field to values $h \in \Big\{{\frac{1}{8},\frac{3}{8},\frac{5}{8},\frac{7}{8}}\Big\}$, which correspond to the average values of the height $z$ in the four columnar ground states [see, for example, \reffig{fig:height}(b)] \cite{Alet2006b}. 

\subsection{Double dimer model}
\label{doubledimermodel}

% Coulomb phase, double dimer model
In the double dimer model, each replica has height field $h^{(\alpha)}$ with identical stiffness $\kappa$, and replicas are coupled by the term $\lambda \bm{\nabla}h^{(1)}\cdot \bm{\nabla} h^{(2)}$, with $\lambda \sim K$. The resulting action for the unsynchronized Coulomb phase may be written
\begin{equation}
\label{eq:DDMCoulomb}
S\sub{DDM} = \int d^{2}\bm{r} \, \left[\frac{\kappa_{+}}{2}\lvert \del h\spr+ \rvert^2 + \frac{\kappa_{-}}{2}\lvert \del h\spr- \rvert^2 \right]
\punc,
\end{equation}
where \(h\spr\pm = h\spr1\pm h\spr2\) and
\begin{equation}
\label{eq:kpm}
\kappa_{\pm} = \frac{1}{2}(\kappa \pm \lambda) \punc.
\end{equation}
Note that in the non-interacting limit, i.e., $J = K = 0$, one has $\lambda = 0$, $\kappa = \kappa_{\infty}$ and
\begin{equation}
\label{eq:kpminf}
\kappa_{\pm,\infty} = \frac{\pi}{2} \punc.
\end{equation}

% Columnar ordering of independent replicas
We now construct field theories that describe phase transitions in the double dimer model. For independent replicas, rotation and translation symmetry constraints apply separately to both $h\spr1$ and $h\spr2$, so each replica has an action given by \refeq{eq:SDMcol}. Therefore, when $K=0$, one expects a columnar-ordering transition with the same critical properties as the single-replica case.

% Synchronization transition action
For the double dimer model with non-zero coupling $K$, we require an action local in both replicas, but now invariant under \emph{simultaneous} $\pi/2$ rotations, and translations, of both replicas. To study the synchronization transition, we focus on the relative height $h\spr-$ [$h\spr+$ remains non-critical] and include additional terms in \refeq{eq:DDMCoulomb}. The constraints on allowed terms are easily derived using results for the single dimer model, and are included in \reftab{table:constraints}. For example, simultaneous translation of dimers by one lattice constant in both replicas maps the height fields $h\spr\alpha(\rv) \rightarrow -h\spr\alpha(\rv - \deltav_{x}) - \frac{1}{4}$, so that the relative height $h\spr-(\rv) \rightarrow -h\spr-(\rv - \deltav_{x})$, which must be a symmetry of the action. In this case, the critical theory is a sine-Gordon model with $p=1$:
\begin{equation}
\label{eq:DDMsync}
S\sub{DDM,sync.} = S\sub{DDM} + \int d^{2}\bm{r} \,V^{(-)}\cos(2\pi h\spr-) \punc,
\end{equation}
where, since the cosine term is forbidden by symmetry constraints when $K=0$, we require $V\spr- \sim K$ to leading order. The constraints imposed by rotation and translation symmetry are identical (see \reftab{table:constraints}), and hence \refeq{eq:DDMsync} remains the correct critical theory for $h\spr-$ in reduced symmetry variations of the double dimer model.

% Ordered phase
From \refeq{eq:rg}, the critical stiffness for the synchronization transition is
\begin{equation}
\label{eq:kappamc}
\kappa_{-,\text{c}} = \frac{\pi}{2} \punc ,
\end{equation}
and ordering occurs when $\kappa_{-} > \kappa_{-,\mathrm{c}}$. The ordered phase is synchronized (antisynchronized) in regions of the phase diagram with negative (positive) coupling $K$, because the cosine term locks $h\spr-=0 \, (\frac{1}{2})$ in order to minimize the action. (The relative height of any synchronized ground state is clearly $h\spr- = 0$.) We identify the (high-temperature) Coulomb phase in the double dimer model with the low-temperature phase of the XY model, in accordance with the duality mapping from integer loops to the XY model \cite{Cardy1996}. Hence, the synchronization transition is a BKT transition but with an inverted temperature axis.

% Critical coupling
To locate the phase boundary at fixed $J/T$, we measure $\kappa_{-}$ as a function of $K/T$ using MC simulations and, from the crossing with its critical value $\kappa_{-,\text{c}}$, identify a critical coupling $(K/T)\sub{c}$. However, in the absence of interactions within replicas ($J=0$) MC simulations are not necessary, because $\kappa_{-,\text{c}}$ precisely coincides with the non-interacting limit ($K=0$) of \refeq{eq:kpminf}. Hence, in this case, the critical coupling $(K/T)\sub{c}=0$, and replicas synchronize for infinitesimal $K<0$ [using \refeq{eq:kpm} with $\lambda \sim K$].

% Columnar ordering of coupled replicas
In our phase diagram, $h\spr-$ is locked in the vicinity of columnar-ordering transitions when $K \neq 0$, and columnar ordering of coupled replicas is thus described by a critical theory in $h\spr+$. Adding to \refeq{eq:DDMCoulomb} the most relevant term consistent with the constraints on $S[h\spr+]$ in \reftab{table:constraints}, one obtains
\begin{equation}
\label{eq:DDMcol}
S\sub{DDM,col.} = S\sub{DDM} + \int d^{2}\bm{r} \,V^{(+)}\cos(4\pi h\spr+) \punc ,
\end{equation}
which is a sine-Gordon model with $p=2$.

The critical stiffness for columnar-ordering of coupled replicas is therefore
\begin{equation}
\label{eq:kappapc}
\kappa_{+,\text{c}} = 2\pi \punc .
\end{equation}
(In principle, $h\spr+$ could lock before $h\spr-$ if $\kappa_{+}>\kappa_{+,\mathrm{c}}$ while $\kappa_{-}<\kappa_{-,\mathrm{c}}$, but we do not observe this.) The ordered phases, for which $\kappa_{+} > \kappa_{+,\mathrm{c}}$, are columnar \& (anti)synchronized. In the columnar \& synchronized phase, for example, where $h\spr - = 0$, the cosine term locks the total height to values $h\spr+ = 2h\spr1 \in \Big\{\frac{1}{4},\frac{3}{4}\Big\}$. This is consistent with average values of the height $z$ for a single dimer model in the columnar phase (see \refsec{singledimermodel}).

\subsection{Honeycomb lattice}
\label{honeycomblattice}

% Introduction
In passing, we consider the double dimer model defined on the honeycomb lattice, which is also bipartite and thus amenable to a height description. As we outline in the following, in the absence of interactions within replicas, i.e., $J=0$, synchronization on the honeycomb lattice occurs at a critical coupling $\left(K/T\right)\sub{c} = 0$, as for the square lattice.

% Stiffness
The Coulomb phase action for the single dimer model on the honeycomb lattice is given by \refeq{eq:SDMCoulomb}, with stiffness fixed to $\kappa = \pi$ by exact calculations \cite{Chalker2017}. This is the same as for the square lattice, \refeq{eq:exactstiffness}, and it follows that the double dimer model is again specified by \refeqs{eq:DDMCoulomb}{eq:kpminf} in the non-interacting limit.

% Critical theory
The constraints on $S[h]$ for the single dimer model are dependent on lattice type: For the honeycomb lattice they become $S[h] =S[h + 1]$ from locality, $S[h] =S[-h]$ from $\pi/3$ rotation symmetry and $S[h] =S[-h - 1/3]$ from translation symmetry (cf. \reftab{table:constraints}). However, for the double dimer model with non-zero coupling $K$, the constraints on $S[h\spr-]$ are unchanged and \refeq{eq:DDMsync} remains the correct critical theory. Hence, as discussed in \refsec{doubledimermodel}, the replicas synchronize for infinitesimal $K<0$.

\begin{figure}
\begin{center}
\includegraphics[width=0.6\columnwidth]{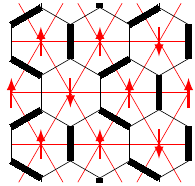}
\caption{Red: a ground state of the triangular lattice Ising antiferromagnet; each plaquette contains a single frustrated bond [parallel spins contributing energy $+\lvert\mathcal{J}\rvert$ in \refeq{eq:TLIAFM}]. Black: corresponding close-packed dimer configuration on the dual (honeycomb) lattice, in which dimers lie across frustrated bonds.}
\label{fig:mapping}
\end{center}
\end{figure}

% SDM <-> TLIAFM
This finding may be interpreted in the context of a simple geometrically frustrated magnet, the triangular lattice Ising antiferromagnet (TLIAFM), which has Hamiltonian
\begin{equation}
\label{eq:TLIAFM}
\ham\sub{TLIAFM} = -\mathcal{J}\sum_{\langle i,j \rangle}\sigma_{i}\sigma_{j} \punc,
\end{equation}
where $\langle i,j \rangle$ denotes nearest-neighbor pairs of sites, $\mathcal{J} < 0$ and $\sigma_{i} = \pm 1$. The TLIAFM has an extensive number of ground states and, as illustrated in \reffig{fig:mapping}, each ground state is in one-to-one correspondence with a close-packed dimer configuration on the honeycomb lattice \cite{Chalker2017}.

% DDM <-> TLIAFM
In the limit $\mathcal{J}/T \rightarrow -\infty$, the double dimer model on the honeycomb lattice is equivalent to a bilayer TLIAFM with Hamiltonian
\begin{equation}
\label{eq:ashkinteller}
\ham = \ham\sub{TLIAFM}\spr{1} + \ham\sub{TLIAFM}\spr{2} + \frac{K}{4}\sum_{\langle i,j \rangle}\sigma_{i}\spr{1}\sigma_{j}\spr{1}\sigma_{i}\spr{2}\sigma_{j}\spr{2} \punc ,
\end{equation}
% all other terms derived from dimer overlap are constants in limit \mathcal{J}/T -> inf
up to additive constants, where the four-spin interaction \cite{Bolton1970, Scott1988, Aplesnin1997} derives from the term that counts overlapping dimers in \refeq{eq:configurationenergy}. Hence, in this limit, spins in both replicas are either all aligned ($\sigma_{i}\spr1 = \sigma_{i}\spr2 \, \forall \, i$) or all antialigned ($\sigma_{i}\spr1 = -\sigma_{i}\spr2 \,\forall \, i$) for infinitesimal $K/T < 0$, according to our height analysis.

In fact, for general $\mathcal{J}$, \refeq{eq:ashkinteller} is the Hamiltonian of the Ashkin--Teller model on the triangular lattice. The phase diagram of this model, obtained using MC simulations in Fig.~7 of \refcite{Lv2011}, includes a BKT critical point at $(\mathcal{J},K) = (-\infty,0)$ and is thus consistent with our conclusion.

Finally, because the honeycomb-lattice dimer loop model is solvable (see \refsec{infinitecoupling}), one may also calculate the stiffnesses $\kappa_{\pm}$ exactly at $(J,K) = (0,+\infty)$, with result $\kappa_{-}=\frac{\pi}{2}=\kappa_{-,\text{c}}$ \cite{Kondev1996}. Hence, this point lies on the synchronization phase boundary. We also observe this feature in the square-lattice phase diagram, \reffig{fig:phasediagram}, where an exact calculation is not possible.

\section{Numerical results}
\label{numericalresults}

In this section we use MC results, obtained using the worm algorithm \cite{Sandvik2006, Wilkins2019}, to map out the phase diagram shown in \reffig{fig:phasediagram} and study the nature of each transition. There are three types of phase boundaries, which we consider in turn: synchronization, columnar ordering, and staggered ordering.

\subsection{Synchronization transitions}
\label{synchronizationtransitions}

% Introduction
In \refsec{doubledimermodel} we identified, when $J/T = 0$, a synchronization transition at infinitesimal coupling between replicas, i.e., $(K/T)\sub{c}=0$. The transition is BKT type, where the Coulomb and synchronized phases correspond to the low- and high-temperature phases of the XY model, respectively. In this section, we first provide MC evidence to support this finding, and then describe how the phase boundary, which divides the Coulomb and (anti)synchronized phases, is located in the case $J/T \neq 0$.

\begin{figure*}
\begin{center}
\includegraphics[width=\textwidth]{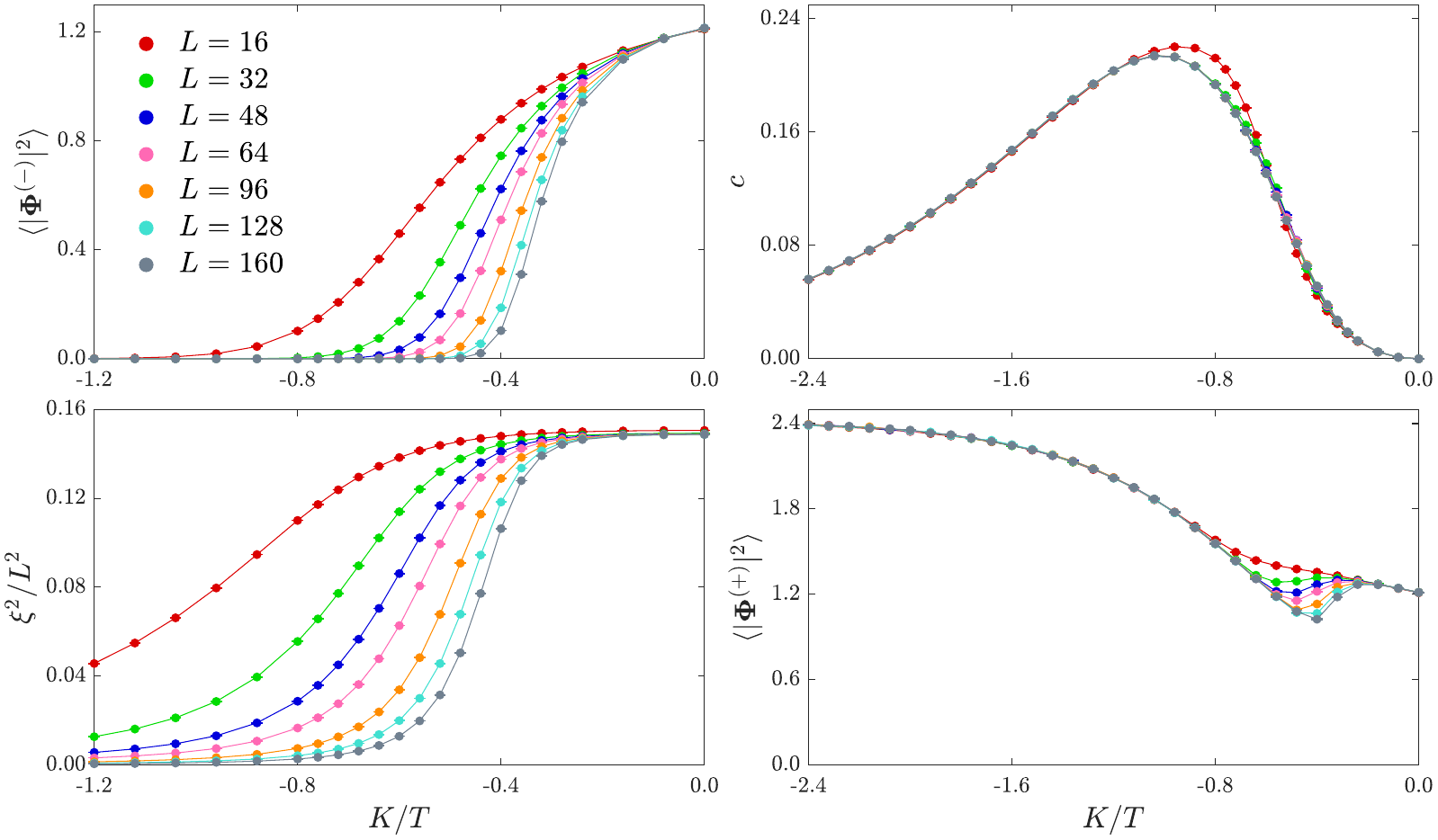}
\caption{Mean-square flux difference $\langle \lvert \bm{\Phi}^{(-)} \rvert ^{2} \rangle$ (top-left panel), heat capacity per site $c$ (top-right panel), square of the normalized confinement length $\xi^{2}/L^{2}$ (bottom-left panel) and mean-square total flux $\langle \lvert \bm{\Phi}^{(+)} \rvert ^{2} \rangle$ (bottom-right panel) vs $K/T$, for the square-lattice double dimer model with $J=0$ and different system sizes $L$. There is a synchronization transition at infinitesimal coupling between replicas, i.e., $(K/T)\sub{c}=0$. The transition is BKT type, where the Coulomb and synchronized phases correspond to the low- and high-temperature phases of the XY model, respectively.}
\label{fig:properties}
\end{center}
\end{figure*}

% MC results for J/T = 0 transition
% Mean-square flux difference
MC data for the synchronization transition when $J/T = 0$ are shown in \reffig{fig:properties}. According to theoretical arguments, the mean-square flux difference $\langle \lvert \bm{\Phi}\spr- \rvert ^{2} \rangle$, shown in the top-left panel, is system-size independent in the Coulomb phase (see \refapp{ddmfluxinthecoulombphase}) and decreases exponentially with $L$ in the synchronized phase. Hence, the extent of the $L$-independent region in $\langle \lvert \bm{\Phi}\spr- \rvert ^{2} \rangle$ provides a rough bound $\lvert (K/T)\sub{c} \rvert \lesssim 0.2$ on the critical coupling. In the thermodynamic limit, however, we expect that $(K/T)\sub{c}$ scales to zero (see \refsec{doubledimermodel}), while $\langle \lvert \bm{\Phi}\spr- \rvert ^{2} \rangle$ jumps discontinuously to zero across the transition. The latter is typical of a BKT transition; for example, in the XY model there is a universal jump in the helicity modulus at the critical point \cite{Nelson1977, Cardy1996}.

% Heat capacity
As expected for a BKT transition, the synchronization transition [at $(K/T)\sub{c}=0$] is not accompanied by a peak in the heat capacity per site $c$, as shown in the top-right panel of \reffig{fig:properties}. Instead, near the transition, theory predicts a non-divergent essential singularity, which is unobservable \cite{Cardy1996, Chaikin2000}. In the XY model, the main feature of the heat capacity per site is a broad peak, which is above the critical temperature and does not diverge with system size. We observe this in the synchronized phase of the double dimer model, i.e., when $K/T < 0$, consistent with the correspondence between the phases in the two models.
% The usual example of an essential singularity is exp(1/z). This is exp(-1/z), so doesn't diverge.

% Confinement length
As discussed in \refsec{coupledreplicas}, the Coulomb and synchronized phases may be distinguished through the monomer confinement criterion. In the bottom-left panel of \reffig{fig:properties}, we show the confinement length $\xi$, defined by
\beq[eq:cl]
\xi^{2} = \frac{\sum_{\Rv} \lvert\Rv\rvert^{2}G\sub{m}(\Rv)}{\sum_{\Rv} G\sub{m}(\Rv)} \punc,
\eeq
which is equivalent to the root-mean-square separation of the test monomers. In the Coulomb phase, where monomers are quasideconfined with \(G\sub{m}(\Rv) \sim \lvert\Rv\rvert^{-\eta}\) (see \refapp{ddmgminthecoulombphase}), the confinement length has asymptotic dependence
\begin{equation}
\xi \sim
\begin{cases}
L & \text{for \phantom{2 < }\(\eta < 2\)} \\
L^{2-\eta/2} & \text{\phantom{for} \(2 < \eta < 4\)} \\ 
L^{0} & \text{\phantom{for} \(4 < \eta\)}
\end{cases}
\end{equation}
(cf. the cubic-lattice case, where fully-deconfined monomers have $\xi\sim L$ independent of stiffness \cite{Wilkins2019}). One also has $\xi \sim L^{0}$ in the synchronized phase, where monomers are confined. In our MC data, the region with $\xi \sim L$ at small $\lvert K/T \rvert$ is thus a signature of a quasideconfined phase. The behavior for large $\lvert K/T \rvert$ is consistent with a confined phase or quasideconfined monomers with $\eta > 4$; we have checked that $G\sub{m}$ decays exponentially in this region (not shown), implying the former. Note that for quasideconfined monomers we observe saturation at $\xi^{2}/L^{2} \simeq 0.15$, which is less than the value expected for fully-deconfined monomers $\xi^{2}/L^{2} \approx 1/6$ [using the result $(L^{2} + 2)/6$ for the mean-square separation of free monomers hopping on an empty lattice].

% Mean-square total flux
In the synchronized phase, where the replicas become strongly correlated, fluctuations in the relative flux $\bm{\Phi}\spr{-}$ are suppressed. However, both replicas remain disordered so fluctuations in the total flux $\bm{\Phi}\spr{+}$ are large in both phases, as shown in the bottom-right panel of \reffig{fig:properties}. In particular, at $K/T = 0$ where the replicas are independent, $\langle \bm{\Phi}^{(1)} \cdot \bm{\Phi}^{(2)} \rangle = \langle \bm{\Phi}^{(1)} \rangle\cdot\langle \bm{\Phi}^{(2)} \rangle = 0$ and $\langle \lvert \bm{\Phi}^{(+)} \rvert ^{2} \rangle = 2 \langle \lvert \bm{\Phi}^{(1)} \rvert ^{2} \rangle$. This is half the value at $K/T \rightarrow -\infty$ for perfectly synchronized replicas, where $\langle \bm{\Phi}^{(1)} \cdot \bm{\Phi}^{(2)} \rangle = \langle \lvert \bm{\Phi}^{(1)} \rvert ^{2} \rangle$ and $\langle \lvert \bm{\Phi}^{(+)} \rvert ^{2} \rangle = 4 \langle \lvert \bm{\Phi}^{(1)} \rvert ^{2} \rangle$.

\begin{figure}
\begin{center}
\includegraphics[width=\columnwidth]{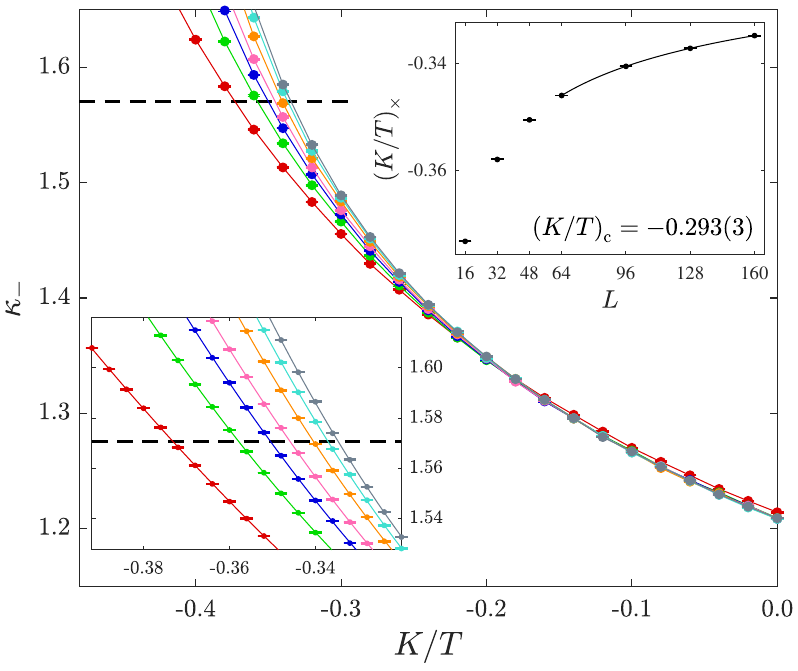}
\caption{Stiffness $\kappa_{-}$ vs $K/T$ for $J/T=0.2$. MC data crosses the critical value $\kappa_{-,\text{c}}=\frac{\pi}{2}$ (dashed line) at the synchronization transition. (Colors indicate different values of $L$ as in \reffig{fig:properties}.) Bottom-left inset: Quadratic fits used to measure a crossing point $(K/T)_{\times}$ for each system size. Top-right inset: $(K/T)_{\times}$ vs system size $L$. The solid line is a fit to \refeq{eq:fss2} for $L \geq 64$, from which a value for the critical coupling $(K/T)\sub{c} = -0.293(3)$ is obtained.}
\label{fig:stiffness}
\end{center}
\end{figure}

% Stiffness
For general $J/T$, we locate the phase boundary between the (anti)synchronized and Coulomb phases as follows. In the Coulomb phase, the mean-square total and relative flux are given by \cite{Alet2006b}
\begin{equation}
\label{eq:DDMCoulombFlux0}
\langle \lvert \bm{\Phi}\spr\pm \rvert ^{2} \rangle = 2\frac{\sum_{n_{1},n_{2} \in \mathbb{Z}}n_{\pm}^{2} e^{-\frac{\kappa_{+}}{2} n_{+}^{2}} e^{-\frac{\kappa_{-}}{2} n_{-}^{2}}}{\sum_{n_{1},n_{2}\in \mathbb{Z}}e^{-\frac{\kappa_{+}}{2} n_{+}^{2}} e^{-\frac{\kappa_{-}}{2} n_{-}^{2}}} \punc ,
\end{equation}
where $n_{\pm} = n_{1} \pm n_{2}$, as derived in \refapp{ddmfluxinthecoulombphase} starting from the continuum theory of \refeq{eq:DDMCoulomb}. In the MC simulations we measure both $\langle \lvert \bm{\Phi}\spr\pm \rvert ^{2} \rangle$, and solve these equations numerically for the stiffnesses $\kappa_{\pm}$ using the Newton--Raphson method \cite{Riley2006}. As shown in \reffig{fig:stiffness} for $J/T = 0.2$, the phase boundary is then located by scanning through $K/T$ until $\kappa_{-}$ crosses its critical value $\kappa_{-,\text{c}} = \frac{\pi}{2}$ [see \refeq{eq:kappamc}].

% Finite-size scaling
To accurately determine the critical coupling, we use quadratic fits to measure a crossing point $(K/T)_{\times}$ for each system size (bottom-left inset of \reffig{fig:stiffness}). For a BKT transition, the appropriate finite-size scaling form is \cite{Bramwell1993, Bramwell1994}
\begin{equation}
\label{eq:fss2}
\left(\frac{K}{T}\right)_{\times} = \left(\frac{K}{T}\right)_{\mathrm{c}} + \frac{A}{\log (L/L_{0})^2} \punc ,
\end{equation}
where $A$ and $L_{0}$ are constants. From our fit for $J/T = 0.2$ (top-right inset of \reffig{fig:stiffness}), we obtain $(K/T)\sub{c}=-0.293(3)$. Ten further critical points located in this way are shown in the phase diagram of \reffig{fig:phasediagram}; this includes transitions into the antisychronized phase at $K>0$ which, notably, all scale onto the line $J/T = 0$.

\subsection{Columnar-ordering transitions}
\label{columnar-orderingtransitions}

% Introduction
Next, we consider transitions into all columnar-ordered phases. For the case of independent replicas, i.e., when $K=0$, columnar ordering at $J < 0$ separates the columnar and Coulomb phases. This transition has been studied in detail by Alet \etal\ in \refcites{Alet2005,Alet2006b} for the single dimer model, where the critical temperature is determined using an order parameter. We first review this approach.

% DRSB
The magnetization \(\Mv\), defined in \refeq{eq:Mdef}, breaks both translation and rotation symmetry in the columnar phase. Denoting by $N_{\mu}$ the number of dimers with orientation $\mu$, a simpler choice of order parameter is the dimer rotation symmetry breaking
\begin{equation}
\label{eq:drsb}
D = \frac{2}{L^{2}}\left\lvert N_{y} - N_{x} \right\rvert \punc ,
\end{equation}
a scalar that is sensitive only to rotation symmetry breaking.

% Mean
This is still sufficient to indicate a columnar-ordering transition: In the Coulomb phase, by symmetry one expects $\langle N_{x} \rangle = \langle N_{y} \rangle$ so that $\langle D \rangle$ is small. In the columnar phase, rotation symmetry is broken and all dimers are either horizontal or vertical. Hence, one expects $\langle D \rangle = 1$ (the total number of dimers is $L^2/2$). Alet \etal\ observe this behavior in Fig.~9 of \refcite{Alet2006b}.

% Binder cumulant scaling
The critical temperature may be determined accurately using the dimer rotation symmetry breaking Binder cumulant
\begin{equation}
\label{eq:K=0bindercumulant}
B_{D} = 1 - \frac{\langle D^{4} \rangle}{3 \langle D^{2} \rangle ^{2}} \punc .
\end{equation}
In the vicinity of the critical point, the $k^{\text{th}}$ moment of the dimer rotation symmetry breaking has scaling form \cite{Wang2006}
\begin{equation}
\langle D^{k} \rangle \sim L^{ka}f_{k}(L/\zeta) \punc,
\end{equation}
where, for a BKT transition, the correlation length diverges as \cite{Cardy1996}
\begin{equation}
\label{eq:BKTcl}
\zeta \sim \exp{\left(bt^{-\frac{1}{2}}\right)} \punc .
\end{equation}
Here, $a$ and $b$ are unknown constants, $f_{k}$ is a universal function, and $t=(T-T\sub{c})/T\sub{c}$ is the reduced temperature. Hence, the Binder cumulant has zero scaling dimension, i.e.,
\begin{equation}
B_{D} \sim g(L/\zeta) \punc ,
\end{equation}
where $g$ is a new universal function, because \refeq{eq:K=0bindercumulant} has equal powers of $D$ in both numerator and denominator.

% Binder cumulant crossing
At the critical temperature $t=0$, the correlation length diverges and, to leading order, the Binder cumulant has no system size dependence. Hence, depending on the finite-size behavior either side of $t=0$, MC data for different system sizes may cross at the critical temperature. This is observed for $B_{D}$ in Fig.~11 of \refcite{Alet2006b}, from which Alet \etal\ report $T\sub{c} = 0.65(1)$ when $J=-1$, but not for the Binder cumulant of $\bm{M}$ \cite{Alet2006b,Papanikolaou2007}.

\begin{figure*}
\begin{center}
\includegraphics[width=\textwidth]{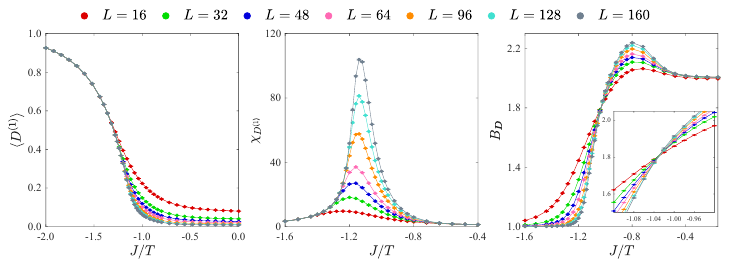}
\caption{Dimer rotation symmetry breaking mean $\langle D\spr{1} \rangle$ (left panel), susceptibility $\chi_{D\spr{1}}$ (middle panel), and Binder cumulant $B_{\bm{D}}$ (right panel) vs $J/T$, for $K/T = -0.2$ and different system sizes $L$. The sharp drop in $\langle D\spr{1} \rangle$, and the corresponding sharp peak in $\chi_{D\spr{1}}$, indicate a phase transition between columnar \& synchronized and synchronized phases. From the crossing in $B_{\bm{D}}$ (right panel, inset), we estimate the critical coupling \((J/T)\sub{c}=-1.03(2)\). This is the generalization of Figs.~9--11 in \refcite{Alet2006b} to the case of coupled replicas.}
\label{fig:CRdrsb}
\end{center}
\end{figure*}

% DRSB mean and susceptibility
We now generalize this method to locate the phase boundary when $K/T \neq 0$, i.e., for coupled replicas. In this case, columnar-ordering transitions separate the (anti)synchronized phases from the columnar \& (anti)synchronized phases. Since translation and rotation symmetry are broken in all columnar-ordered phases, we again expect a sharp drop in the mean dimer rotation symmetry breaking $\langle D\spr\alpha \rangle$ for both replicas, as well as a peak in the corresponding susceptibility
\begin{equation}
\chi_{D^{(\alpha)}} = L^{2}\left(\langle \left[D^{(\alpha)}\right]^{2} \rangle - \langle D^{(\alpha)} \rangle^{2}\right) \punc ,
\end{equation}
in the vicinity of a transition. This is shown in \reffig{fig:CRdrsb} (left and middle panels) for the transition at $K/T = -0.2$, between the columnar \& synchronized and synchronized phases.

% DRSB Binder cumulant
To measure the critical coupling, we have analyzed the Binder cumulant $B_{D\spr\alpha}$ of \refeq{eq:K=0bindercumulant} for both replicas, but no longer observe a distinct crossing point in the MC data when $K \neq 0$. Instead, we define the two-component vector $\bm{D} = (D\spr1,D\spr2)$, with corresponding Binder cumulant
\begin{equation}
\label{eq:Kneq0bindercumulant}
B_{\bm{D}} = \frac{\langle \lvert\bm{D}\rvert^{4} \rangle}{\langle \lvert\bm{D}\rvert^{2} \rangle^{2}} \punc ,
\end{equation}
which is shown in \reffig{fig:CRdrsb} (right panel). Up to normalization, this is equivalent to \refeq{eq:K=0bindercumulant} in the limits $K=0$ and $K/T \rightarrow -\infty$. Deep within the columnar \& (anti)synchronized phases, the probability distribution for $\bm{D}$ is sharply peaked at $D\spr1 = D\spr2 = 1$. Then $\langle \lvert \bm{D} \rvert^{4} \rangle = \langle \lvert \bm{D} \rvert ^{2} \rangle^{2}$ so $B_{\bm{D}}$ saturates to unity. In the (anti)synchronized phases, $D^{(\alpha)}$ follows a half-normal distribution for which $\langle \left[D^{(\alpha)}\right]^{4} \rangle = 3\langle \left[D^{(\alpha)}\right]^{2}\rangle^{2}$. The limiting value depends on $K/T$ through the correlator $\langle \left[D\spr1 D\spr2\right]^{2}\rangle$, which can only be calculated at $K=0$ and $K/T \rightarrow -\infty$, where $B_{\bm{D}} = 2$ and $3$, respectively.

% Phase diagram
As shown in \reffig{fig:CRdrsb} (right panel, inset), MC data for this Binder cumulant, $B_{\bm{D}}$, does exhibit a crossing point when $K\neq0$. From this, we estimate a critical coupling $\left(J/T\right)\sub{c} = -1.03(2)$ at $K/T = -0.2$. Our phase diagram, \reffig{fig:phasediagram}, includes this point along with ten others that have been obtained in the same way, but using only system sizes $L=32$ and $L=48$.

% Critical properties
Columnar ordering in the limits $K = 0$, studied in \refcites{Alet2005, Alet2006b}, and $K/T \rightarrow -\infty$, equivalent to columnar ordering of a single dimer model with $J\sub{eff} = 2J$ (see \refsec{infinitecoupling}), is known to be a BKT transition with an inverted temperature axis. We expect the whole phase boundary to share the same critical properties as these points.

\begin{figure}
\begin{center}
\includegraphics[width=\columnwidth]{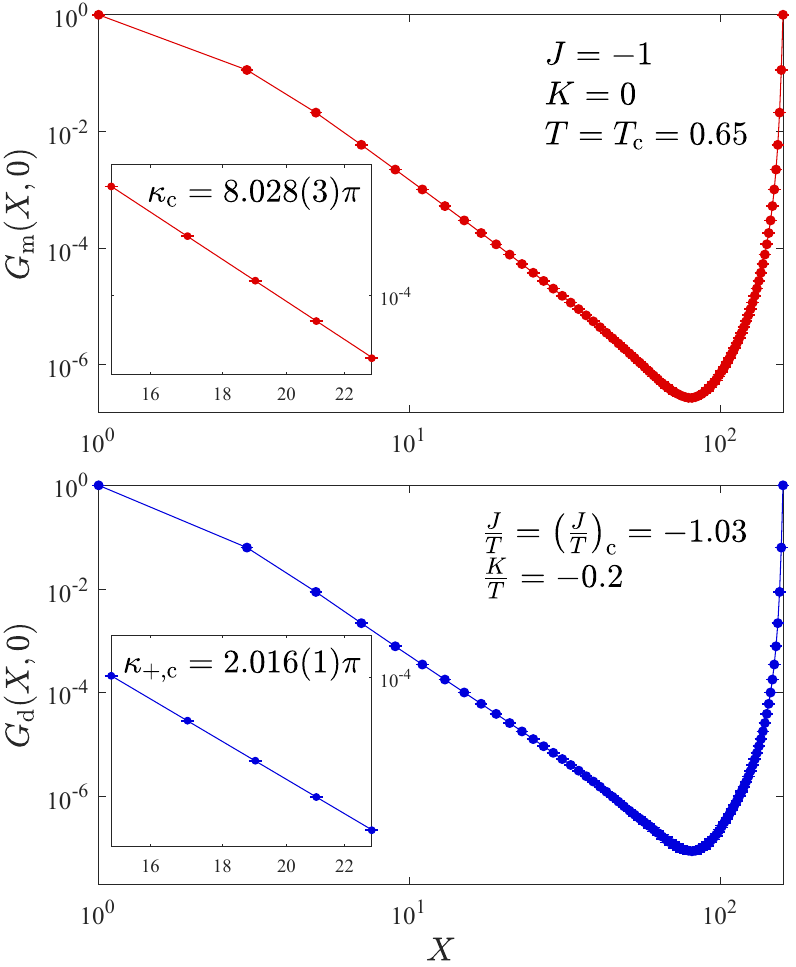}
\caption{Top panel: Log--log plot of monomer distribution function $G\sub{m}(X,0)$ vs monomer separation $X$ at the columnar-ordering transition for independent replicas. Bottom panel: Log--log plot of double monomer distribution function $G\sub{d}(X,0)$ vs $X$ at a columnar-ordering transition for coupled replicas. In each case, the system size is $L=160$. Insets: Solid lines are fits to \refeqand{eq:gmfit}{eq:gdfit} for $15 \geq X \geq 23$, from which values for the critical stiffness $\kappa\sub{c} = 8.028(3)\pi$ and $\kappa_{+,\textrm{c}}=2.016(1)\pi$ are obtained, respectively.}
\label{fig:monomers}
\end{center}
\end{figure}

% Gm general
We now use the field theory and RG analysis of \refsec{fieldtheoriesandcriticalproperties} to verify our results. In \reffig{fig:monomers} (top panel), we measure the monomer distribution function $G\sub{m}(\bm{R})$ at the columnar-ordering transition for independent replicas ($J=-1$, $K=0$ and $T=T\sub{c} = 0.65$), counting only monomers on the same row, i.e., $\bm{R} = (X, 0)$. Each MC simulation can only construct $G\sub{m}$ up to an arbitrary multiplicative constant, so we fix $G\sub{m}(1,0) = 1$.

% Gm fit
The Coulomb phase monomer distribution function has asymptotic form
\begin{equation}
\label{eq:gmfit}
G\sub{m}(X,0) \sim X ^{-\kappa/2\pi} \punc ,
\end{equation}
which is derived in \refapp{sdmgminthecoulombphase} starting from the continuum theory of \refeq{eq:SDMCoulomb}. Due to periodic boundaries, $G\sub{m}(X,0)$ is symmetric around $X = L/2$ in the MC simulations, hence the algebraic decay is cut off and \refeq{eq:gmfit} is only valid for $1 \ll X \ll L/2$. A fit to \refeq{eq:gmfit} over a suitable range in the inset yields an estimate for the critical stiffness $\kappa\sub{c} = 8.028(3)\pi$, which is comparable with the RG prediction $\kappa\sub{c} = 8\pi$ of \refeq{eq:kappac} [the discrepancy perhaps arises due to the uncertainty in $T\sub{c}$ and finite-size effects in $G\sub{m}(X,0)$]. Alet \etal\ instead measure the flux and invert \refeq{eq:fluxCoulomb} to obtain the stiffness (see Fig.~31 of \refcite{Alet2006b}).

% Gd
In the case of coupled replicas, one requires the asymptotic form of $G\sub{m}(\bm{R})$ in the (anti)synchronized phases, which is less straightforward. Instead, it is convenient to consider the double monomer distribution function $G\sub{d}(\bm{R})$ (see \refsec{coupledreplicas}) which, as derived in \refapp{ddmgdinthe(anti)synchronizedphases}, has asymptotic form \footnote{The height description of the columnar-ordering transitions necessarily implies a discontinuity in the phase boundary at ${K=0^{\pm}}$. To see this, compare ${G\sub{d}}$ on the phase boundary at ${K=0}$ and ${K=0^{\pm}}$: the former is equivalent to ${G\sub{m}^{2}}$, where ${G\sub{m}}$ is given by \refeq{eq:heightgm} and ${\kappa=\kappa\sub{c}=8\pi}$, hence ${G\sub{d}\sim \lvert\Rv\rvert^{-8}}$. The latter, however, is given by \refeq{eq:heightgd} with ${\kappa_{+} =\kappa\sub{+,c} =2\pi}$, i.e., ${G\sub{d} \sim \lvert \bm{R} \rvert ^{-4}}$. We have checked that this discontinuity is small and, indeed, it is not resolved by our Binder cumulant method. Such an effect, though, can be seen in the phase diagram of \refcite{Otsuka2009}.}
\begin{equation}
\label{eq:gdfit}
G\sub{d}(X,0) \sim X ^{-2\kappa_{+}/\pi} \punc .
\end{equation}
In \reffig{fig:monomers} (bottom panel), we show $G\sub{d}(\bm{R})$ for the columnar-ordering transition at $K/T = -0.2$ and $J/T=(J/T)\sub{c} = -1.03$. In the inset, a fit to \refeq{eq:gdfit} over a suitable range gives $\kappa_{+,\text{c}}=2.016(1)\pi$, which is close to the expected value $\kappa_{+,\mathrm{c}}=2\pi$ of \refeq{eq:kappapc}.

\subsection{Staggered-ordering transitions}
\label{staggeredorderingtransitions}

\begin{figure}
\begin{center}
\includegraphics[width=\columnwidth]{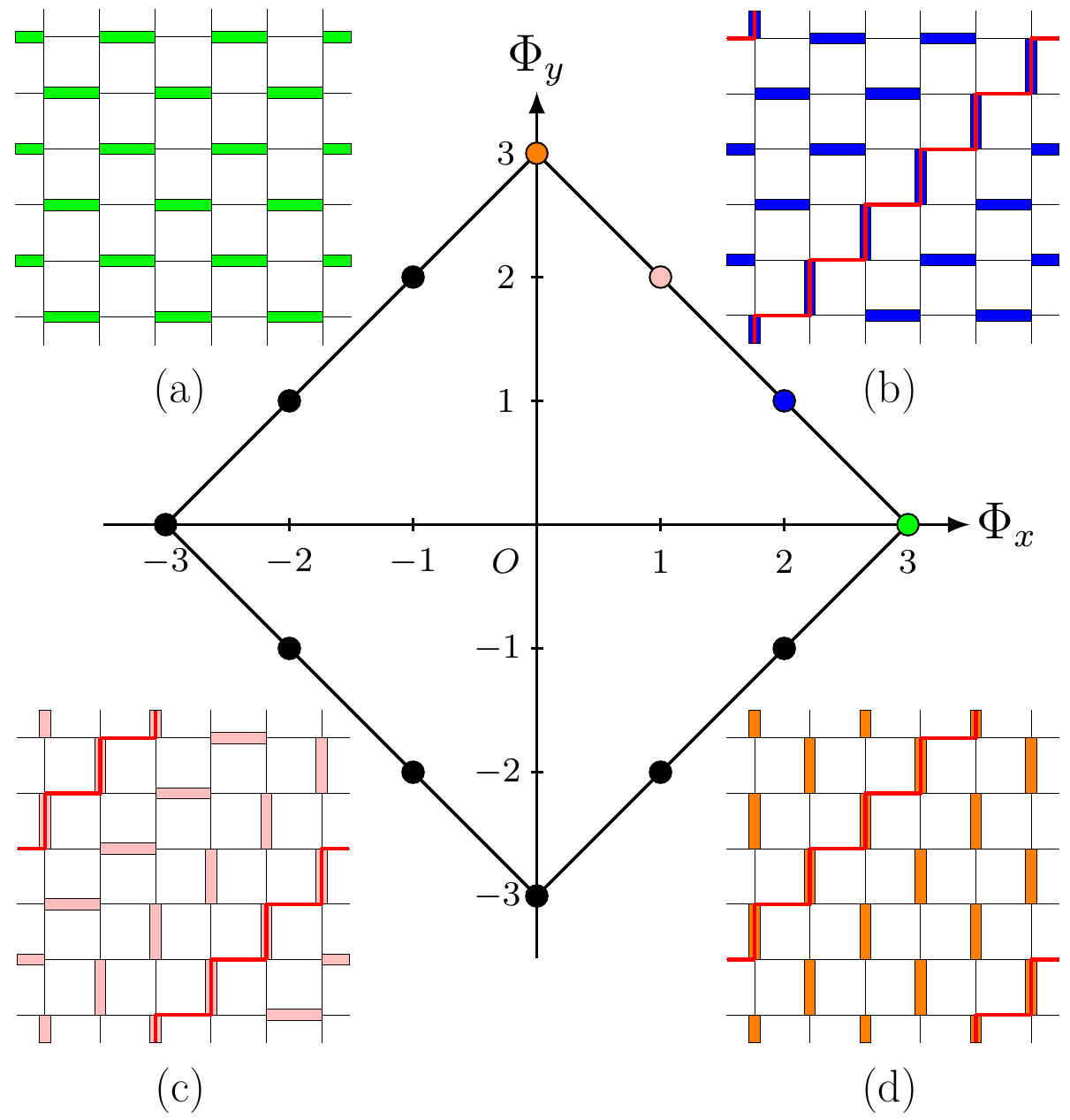}
\caption{Examples of staggered ground states for a single dimer model on a $6 \times 6$ lattice with periodic boundaries. Ground states (a) and (d) have all dimers horizontal and vertical, respectively. Ground states (b), (c) and (d) are related to (a), (b) and (c), respectively, by translation of dimers around red `staircases'. Center: Ground state manifold in flux space, described by the equation $\lvert\Phi_{x}\rvert + \lvert\Phi_{y}\rvert = L/2$; a dot with flux $\bm\Phi = (\Phi_{x}, \Phi_{y})$ corresponds to ${}^{L/2}C_{\lvert \Phi_{x} \rvert}$ degenerate ground states (see text for explanation). Colored dots correspond to the positions of ground states (a)--(d).}
\label{fig:fluxgs}
\end{center}
\end{figure}

% Staggered ground states
We begin by describing the nature of the staggered phase in the single dimer model. The simplest staggered ground states (which contain no parallel pairs of dimers) have all dimers horizontal, such as \reffig{fig:fluxgs}(a), or vertical, such as \reffig{fig:fluxgs}(d). More complicated ground states are obtained by shifting dimers along diagonal loops, or `staircases', that span the periodic boundaries. For example, \reffig{fig:fluxgs}(b) is a staggered ground state related to \reffig{fig:fluxgs}(a) by translation of dimers around the red staircase. Translation of dimers around further staircases results in \reffig{fig:fluxgs}(c), and then \reffig{fig:fluxgs}(d).

% Ground state manifold in flux space
From \refeq{eq:fluxdef}, the ground state in \reffig{fig:fluxgs}(a) has flux $\bm\Phi = (L/2,0)$. Introduction of each staircase reduces (increases) the number of horizontal (vertical) dimers by $L$, resulting in a flux change $\Delta \bm\Phi = (-1,1)$. Consequently, the subset of ground states in Figs.~\ref{fig:fluxgs}(a)--(d) occupy the line $\Phi_{x} +\Phi_{y} = L/2$ in flux space, as illustrated by the center of \reffig{fig:fluxgs}. More generally, the full ground state manifold is given by the equation
\begin{equation}
\label{eq:groundstatemanifold}
\lvert\Phi_{x}\rvert + \lvert\Phi_{y}\rvert = \frac{L}{2} \punc .
\end{equation}
This simple representation of the staggered ground states is specific to two dimensions, and cannot be generalized to the cubic lattice.

% Degeneracy
There is only one ground state, shown in \reffig{fig:fluxgs}(a), with flux $\bm{\Phi} = (L/2,0)$. To construct configurations with $\Phi_{y} > 0$ [for example Figs.~\ref{fig:fluxgs}(b)--(d)] one must insert $\Phi_{y}$ staircases into $L/2$ slots, for which the number of choices is given by the binomial coefficient ${}^{L/2}C_{\Phi_{y}}$. In general, the degeneracy of a staggered ground state with flux $\bm{\Phi} = (\Phi_{x}, \Phi_{y})$ is ${}^{L/2}C_{\lvert\Phi_{y} \rvert}$ [or equivalently, by \refeq{eq:groundstatemanifold}, ${}^{L/2}C_{\lvert\Phi_{x} \rvert}$].

% Entropy
Using this binomial distribution, one may calculate observables deep within the staggered phase. For example, the total number of ground states is
\begin{align}
N &= 4\left(\sum_{\Phi_{x}=0}^{L/2} \ {}^{L/2}C_{\lvert \Phi_{x} \rvert} - 1\right) \\ 
\label{eq:entropy} &= 4(2^{L/2} - 1),
\end{align}
which corresponds to a subextensive entropy $\log{N} \approx \frac{L}{2}\log{2}$. We also infer that, since the quantity \(\lvert\Phi_x\rvert/L\) is distributed around \(1/4\) with standard deviation \(\propto L^{-1/2}\), the flux takes one of the four values \(\Phiv/L = (\pm \frac{1}{4}, \pm \frac{1}{4})\) in the thermodynamic limit, thus spontaneously breaking rotation and translation symmetries. At finite \(J\) there are fluctuations out of these extremal states, but the symmetry-breaking transition remains.

For the double dimer model, the above discussion allows one to write down the partition function exactly, as a function of $K/T$, in the limit $J/T \rightarrow \infty$. For example, for both replicas consider only ground states in the first quadrant of \reffig{fig:fluxgs}, i.e., \(0 \le \Phi_{x,y}\le L/2\). A staircase can be covered by either horizontal or vertical dimers within each replica, and so, including a field $\bm{h}$ that couples to the flux difference $\Phiv\spr-$, has four possible Boltzmann weights: $\exp(-KL/T)$ when both replicas have the same orientation (i.e., both horizontal or vertical), and $\exp[\pm\bm{h}\cdot (1,-1)]$ when both replicas have different orientations. Since, in total, there are $L/2$ staircases, the contribution of these ground states to the partition function is
\begin{equation}
Z_{11} = 2^{L/2}\left[e^{-KL/T} + \cosh{(h_{x} - h_{y}})\right]^{L/2} \punc .
\end{equation}

Because $Z_{11}$ contains all configurations with maximal overlap, we expect that, for $K=0^{-}$, the full partition function asymptotically approaches $Z_{11}$ in the thermodynamic limit. By taking suitable derivatives with respect to $\bm{h}$, one finds that the flux difference $\Phiv\spr-$ is distributed around $\bm{0}$ with variance $\approx Le^{-\lvert K\rvert L/T}$. Hence, as illustrated in \reffig{fig:phasediagram}, in the staggered phase infinitesimal negative coupling is sufficient to synchronize the two replicas. Similarly, for $K>0$, one expects $\bm{\Phi}\spr1/L=(\pm\frac{1}{4},\pm\frac{1}{4})$ in the thermodynamic limit, $\bm{\Phi}\spr2=-\bm{\Phi}\spr1$ to minimize overlap, and hence $\bm{\Phi}\spr+=0$ (we note that there are also, for example, configurations with $\bm{\Phi}\spr-=0$ and zero overlap, but their degeneracy is less by a factor exponentially small in $L$).

% Phase transitions
We now use MC results to examine transitions into all staggered-ordered phases. To begin, we focus on the case $K=0$, where staggered ordering at $J > 0$ separates the staggered and Coulomb phases. One expects the same critical properties as for the single-replica case so, for simplicity, we consider a single dimer model with $J = +1$ and vary the temperature. 

\begin{figure*}
\begin{center}
\includegraphics[width=\textwidth]{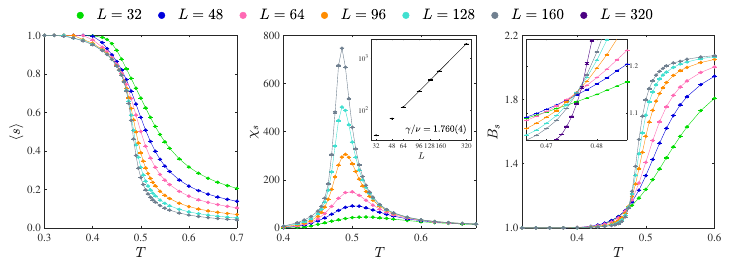}
\caption{Staggered order parameter mean $\langle s\rangle$ (left panel), susceptibility $\chi_{s}$ (middle panel), and Binder cumulant $B_{s}$ (right panel) vs temperature $T$, for the square-lattice dimer model (two independent replicas, $K=0$) with $J = +1$ and different system sizes $L$. The sharp drop in $\langle s \rangle$, and the corresponding sharp peak in $\chi_{s}$, indicate a phase transition between staggered and Coulomb phases. Right panel, inset: From the crossing in $B_{s}$, we estimate the critical temperature \(T\sub{c}=0.477(2)\). Middle-panel, inset: Log--log plot of $\chi_{s}$, evaluated at the critical temperature $T\sub{c}=0.477$, versus system size $L$. The solid line is a fit to \refeq{eq:chisscaling} for $L \geq 64$, from which a value $\gamma/\nu=1.760(4)$ is obtained.}
\label{fig:IRsop}
\end{center}
\end{figure*}

% Order parameter
By analogy with the columnar-ordering transitions (cf. \reffig{fig:CRdrsb}), we use the staggered order parameter \cite{Alet2016}
\begin{equation}
s = \frac{2}{L}\left(\lvert \Phi_{x} \rvert + \lvert \Phi_{y} \rvert\right)
\end{equation}
to determine the critical temperature. At low temperatures, deep within the staggered phase, one has $\langle s \rangle = 1$ by definition of the ground state manifold, \refeq{eq:groundstatemanifold}, whereas in the Coulomb phase $\langle s \rangle$ is small because the flux distribution \(P(\Phiv)\) is peaked at \(\Phiv=\zerov\) with width \(\sim L^0\) [see \refeq{eq:probflux}]. Between these regimes, the sharp drop in $\langle s \rangle$ and peak in the corresponding susceptibility 
\begin{equation}
\chi_{s} = L^{2}\left(\langle s^{2} \rangle - \langle s \rangle^{2}\right) \punc ,
\end{equation}
shown in \reffig{fig:IRsop} (left and middle panels), are characteristic of a phase transition.

% Binder cumulant
In \reffig{fig:IRsop} (right panel), we obtain the critical temperature from the crossing point in the staggered order parameter Binder cumulant \cite{Alet2016}
\begin{equation}
B_{s} = \frac{\langle s ^{4} \rangle}{\langle s^{2} \rangle ^{2}} \punc .
\end{equation}
% Dropping the usual normalization, because don't obtain usual high-temperature limit B_s = 0 (i.e. s is not Gaussian)
Our estimate, $T\sub{c} = 0.477(2)$, is close to existing results \(T\sub{c}=0.449(1)\) and \(T\sub{c}=0.51\) of \refcites{Alet2005, Otsuka2009}, respectively [see also \refcite{Castelnovo2007}, which reports \(T\sub{c}=0.72(5)\)].

% Critical properties
The absence of relevant cosine terms in the action for $J>0$ implies that staggered ordering does not occur through a BKT transition, and is consistent with either a first-order transition, as suggested by Castelnovo \etal\ \cite{Castelnovo2007}, or a standard Landau-type ordering transition. Our MC data suggest that the transition is in fact continuous: $B_{s}$ has a crossing point, while the heat capacity per site $c$ (not shown) does not diverge strongly with system size (i.e., not $\sim L^{2}$).

At the critical point for a continuous transition, the susceptibility has finite-size scaling form \cite{Cardy1996}
\begin{equation}
\label{eq:chisscaling}
\chi_{s} \sim L^{\gamma/\nu} \punc ,
\end{equation}
where $\gamma$ and $\nu$ are the susceptibility and correlation-length exponents, respectively. A fit to this form in \reffig{fig:IRsop} (middle panel, inset) yields a rough estimate $\gamma/\nu = 1.760(4)$, where the error reflects the quality of the fit, but ignores uncertainty in $T\sub{c}$ and higher-order corrections to \refeq{eq:chisscaling}. This is close to $\gamma/\nu$ in the Ising ($\gamma=7/4$, $\nu=1$), 3-state Potts ($\gamma = 13/9$, $\nu = 5/6$) and Ashkin--Teller ($\gamma = 7/6$, $\nu = 2/3$) 2D universality classes \cite{Baxter1982}. Based on the four values \(\Phiv/L = (\pm \frac{1}{4}, \pm \frac{1}{4})\) taken by the flux deep within the ordered phase, a na\"ive Landau theory would be that of the \(4\)-state clock model, which is equivalent to two uncoupled Ising models \cite{Suzuki1967} and thus supports the Ising universality class. Confirmation of this would require a more detailed analysis, which is beyond the scope of this work.

In the height picture, the transition occurs when the stiffness $\kappa=0$ in the Gaussian action, \refeq{eq:SDMCoulomb}. For this reason, Otsuka \cite{Otsuka2009} and Alet \cite{Alet2016} have made the connection with the quantum spin-$\frac{1}{2}$ XXZ chain, and spin ice subjected to uniaxial pressure \cite{Jaubert2010}, for which all terms in the action vanish to infinite order at the critical point (by symmetry for the XXZ chain; `accidentally' for spin ice under pressure). Such infinite-order multicritical points \cite{Benguigui1977} exhibit both first-order and continuous features. Since we do not observe the former, our results suggest that higher-order terms do not vanish in the dimer model, i.e., $P(\bm{\Phi})$ is not flat [see \refeq{eq:probflux}], at $T\sub{c}$.

% Double dimer model
To locate the full phase boundary our approach is straightforwardly extended to the case of coupled replicas, using crossing points in the Binder cumulant
\begin{equation}
B_{\bm{s}} = \frac{\langle \lvert\bm{s}\rvert^{4} \rangle}{\langle \lvert\bm{s}\rvert^{2} \rangle^{2}} \punc ,
\end{equation}
where $\bm{s} = (s^{(1)},s^{(2)})$. Eleven such points are included in our phase diagram, \reffig{fig:phasediagram}, obtained for system sizes $L=64$ and $L=96$. We again infer the critical properties of the whole phase boundary from the limits $K=0$ and $K/T \rightarrow -\infty$.

\section{Conclusions}
\label{conclusions}

Our central result is the phase diagram of the classical double dimer model on the square lattice, shown in \reffig{fig:phasediagram}. As on the cubic lattice, we find a synchronization phase transition at which fluctuations between the two replicas become more strongly correlated, with signatures in the variance of the relative flux and in the monomer distribution function, but no symmetry breaking. The critical properties at this transition are of the BKT type, as expected for such a transition in 2D.

In addition, we find an antisynchronized phase, which was not observed on the cubic lattice, where overlaps between the two replicas are reduced. Our numerical results indicate that the phase boundary with the Coulomb phase runs along the line \(J/T = 0\) for positive \(K/T\) (except possibly close to \(K=0\), where the finite-size scaling becomes more difficult), as has previously been conjectured \cite{Raghavan1997}.

Remarkably, and in contrast with the 3D case, we find that these three phases meet at the noninteracting point \(J = K = 0\), implying that an infinitesimal coupling between replicas is sufficient to drive the synchronization transition. This conclusion is supported both by our numerical results and by theoretical considerations based on a height field theory.

In forthcoming work \cite{BosonizationPaper} we will apply bosonization to the transfer-matrix solution of the dimer model \cite{Lieb1967}. This provides an alternative perspective on the fact that the synchronization transition is at infinitesimal coupling, because it can be understood as a pairing transition for fermions at zero temperature in 1D. It also allows one to predict the asymptotic form of the phase boundary exactly, based on perturbation theory in terms of the couplings.

While we have focused here on the case of the square lattice, we have also shown that the double dimer model on the honeycomb lattice similarly synchronizes for infinitesimal attractive coupling. Since this model is equivalent to a bilayer triangular lattice Ising antiferromagnet \cite{Chalker2017}, this suggests a possible experimental realization. The synchronization transition would manifest as a highly unusual thermal phase transition between distinct correlated paramagnets, though establishing clear experimental signatures would likely be challenging.

A natural extension of the system studied here would involve multiple replicas. With sufficiently strong coupling between adjacent pairs, this could be interpreted as a trajectory either of the classical dimer model imbued with dynamics or of a quantum dimer model in imaginary time. Alternatively, coupling one replica to \(n\) others and taking the limit \(n\rightarrow 0\) \cite{Cardy1996} provides a way to introduce a quenched disorder potential on the links of the single dimer model.

\acknowledgments{The simulations used resources provided by the University of Nottingham High-Performance Computing Service. We are grateful to F. Alet and J. P. Garrahan for helpful discussions.}

\appendix

\section{Calculation of observables using field theories}
\label{calculationofobservablesusingfieldtheories}

In this Appendix, we calculate various observables in the single dimer model (SDM) and double dimer model (DDM) using the continuum theories introduced in \refsec{fieldtheoriesandcriticalproperties}.

\subsection{SDM flux in the Coulomb phase}
\label{sdmfluxinthecoulombphase}

% Definition
We first calculate flux moments in the SDM Coulomb phase; a similar version of this derivation can be found in \refcites{Alet2006b,Tang2011}. The SDM Coulomb phase action is given by
\beq
\label{eq:SDMCoulomb2}
S\sub{SDM} = \int d^{2}\bm{r} \, \frac{\kappa}{2}\lvert \bm{B} \rvert^2 \punc,
\eeq
and the probability associated with magnetic field $\bm{B}(\rv)$ is $P[\bm{B}] = e^{-S\sub{SDM}[\bm{B}]}/Z$, where $Z$ is the partition function.

% Fourier expansion
We write the magnetic field as a Fourier series
\begin{equation}
\bm{B}(\rv) = \frac{1}{L}\sum_{\bm{k}}e^{-i\bm{k}\cdot\rv}\tilde{\bm{B}}(\bm{k}) \punc ,
\end{equation}
with Fourier coefficients
\begin{equation}
\tilde{\bm{B}}(\bm{k}) = \frac{1}{L}\int d^{2}\rv \, e^{i\bm{k}\cdot\rv}\bm{B}(\rv) \punc .
\end{equation}
Formally, the coarse-graining procedure is defined by
\begin{equation}
B_{\mu}(\rv) = \sum_{\bm{r}'}B_{\bm{r}',\mu}\mathcal{K}_{w}(\bm{r} - \bm{r}') \punc ,
\end{equation}
where $\mathcal{K}_{w}$ is a coarse-graining kernel with width $w$ on the order of a few lattice spacings, and normalization
\begin{equation}
\int d^{2}\rv \, \mathcal{K}_{w}(\rv - \bm{r}') = 1.
\end{equation}
Hence, the Fourier coefficient
\begin{align}
\tilde{B_{\mu}}(\bm{0}) &= \frac{1}{L}\int d^{2}\rv \, B_{\mu}(\rv) \\
&= \frac{1}{L} \sum_{\bm{r}'}B_{\bm{r}',\mu}\int d^{2}\rv \, \mathcal{K}_{w}(\rv - \bm{r}')\\
&= \Phi_{\mu} \punc,
\end{align}
where $\Phi_{\mu}$ is the flux defined by \refeq{eq:fluxdef}.
% But other Fourier components are continuous

In terms of the Fourier coefficients, the action becomes
\begin{equation}
S\sub{SDM}[\bm{B}] = \sum_{\bm{k}} \frac{\kappa}{2} \lvert\tilde{\bm{B}}(\bm{k}) \rvert^{2} \punc .
\end{equation}
The probability of flux $\bm{\Phi}$ is obtained by integrating out all other Fourier modes with $\bm{k}\neq \bm{0}$, so
\begin{align}
P(\bm{\Phi}) &= \int \prod_{\bm{k} \neq \bm{0}}d^{2}\tilde{\bm{B}}(\bm{k}) \, P[\bm{B}] \\
\label{eq:probflux}
&= \frac{e^{-\frac{\kappa}{2}\bm{\Phi}^{2}}}{\sum_{\bm{\Phi}}e^{-\frac{\kappa}{2}\bm{\Phi}^{2}}} \punc ,
\end{align}
where $\Phi_{\mu}$ is integer valued. As expected, the mean flux vanishes while the mean-square flux is given by \cite{Alet2006b}
\begin{equation}
\label{eq:fluxCoulomb}
\langle \lvert \bm{\Phi} \rvert ^{2} \rangle = 2\frac{\sum_{n \in \mathbb{Z}} n^{2} e^{-\frac{\kappa}{2} n^{2}}}{\sum_{n\in\mathbb{Z}}e^{-\frac{\kappa}{2} n^{2}}} \punc .
\end{equation}
Unlike in the case of the cubic lattice \cite{Alet2006}, the discreteness of the flux is important in two dimensions and the sum over flux sectors cannot be converted into an integral.

\subsection{DDM flux in the Coulomb phase}
\label{ddmfluxinthecoulombphase}

% Generalization to double dimer model
The generalization to flux moments in the DDM Coulomb phase is straightforward. The DDM Coulomb phase action is
\begin{equation}
\label{eq:DDMCoulombB}
S\sub{DDM} = \int d^{2}\bm{r} \, \left[\frac{\kappa_{+}}{2}\lvert \bm{B}\spr+ \rvert^2 + \frac{\kappa_{-}}{2}\lvert \bm{B}\spr- \rvert^2 \right]
\punc,
\end{equation}
and the probability associated with magnetic fields $\bm{B}\spr\pm(\rv)$ is $P[\bm{B}\spr+, \bm{B}\spr-] = e^{-S\sub{DDM}[\bm{B}\spr+, \bm{B}\spr-]}/Z$, where $Z$ is the partition function.

After Fourier expansion of $\bm{B}\spr\pm(\rv)$ in terms of Fourier coefficients $\tilde{\bm{B}}\spr\pm(\bm{k})$, where 
$\tilde{\bm{B}}\spr\pm(\bm{0}) = \bm{\Phi}\spr\pm$, the action becomes
\begin{equation}
S\sub{DDM} = \sum_{\bm{k}} \left[ \frac{\kappa_{+}}{2} \lvert\tilde{\bm{B}}\spr+(\bm{k}) \rvert^{2} + \frac{\kappa_{-}}{2} \lvert\tilde{\bm{B}}\spr-(\bm{k}) \rvert^{2} \right] \punc .
\end{equation}
The probability of fluxes $\bm{\Phi}\spr\pm$ is obtained by integrating out all other Fourier modes with $\bm{k}\neq \bm{0}$, so
\begin{equation}
P(\bm{\Phi}\spr+, \bm{\Phi}\spr-) = \frac{e^{-\frac{\kappa_{+}}{2}\lvert\bm{\Phi}\spr+\rvert^{2}}e^{-\frac{\kappa_{-}}{2}\lvert\bm{\Phi}\spr-\rvert^{2}}}{\sum_{\bm{\Phi}\spr1,\bm{\Phi}\spr2}e^{-\frac{\kappa_{+}}{2}\lvert\bm{\Phi}\spr+\rvert^{2}}e^{-\frac{\kappa_{-}}{2}\lvert\bm{\Phi}\spr-\rvert^{2}}} \punc ,
\end{equation}
where $\Phi_{\mu}\spr{1,2}$ are integer valued (we avoid summing over $\Phi_{\mu}\spr\pm = \Phi_{\mu}\spr1 \pm \Phi_{\mu}\spr2$, which are instead pairs of integers with the same parity). Again, the mean flux vanishes while the mean-square total and relative flux are given by
\begin{equation}
\label{eq:DDMCoulombFlux}
\langle \lvert \bm{\Phi}\spr\pm \rvert ^{2} \rangle = 2\frac{\sum_{n_{1},n_{2} \in \mathbb{Z}}n_{\pm}^{2} e^{-\frac{\kappa_{+}}{2} n_{+}^{2}} e^{-\frac{\kappa_{-}}{2} n_{-}^{2}}}{\sum_{n_{1},n_{2}\in \mathbb{Z}}e^{-\frac{\kappa_{+}}{2} n_{+}^{2}} e^{-\frac{\kappa_{-}}{2} n_{-}^{2}}} \punc ,
\end{equation}
where $n_{\pm} = n_{1} \pm n_{2}$.

\subsection{SDM $G\sub{m}(\bm{R})$ in the Coulomb phase}
\label{sdmgminthecoulombphase}

Next, we calculate the monomer distribution function in the SDM Coulomb phase (see also \refcites{Chalker2017,Tang2011}). In the continuum description, this is given by
\begin{equation}
\label{eq:gmaction}
G\sub{m}(\bm{R}) = \frac{1}{Z}\int \mathcal{D}\bm{B}(\rv) \, e^{-S\sub{SDM}[\bm{B}]} \punc ,
\end{equation}
where $Z$ is the partition function in the close-packed case, $S\sub{SDM}$ is given by \refeq{eq:SDMCoulomb2} and $\bm{B}$ is now the magnetic field in the presence of a pair of test monomers, i.e., $\del \cdot \bm{B} = Q(\rv)$ with $Q(\rv) = \mathcal{K}_{w}(\rv - \bm{r}_{+}) - \mathcal{K}_{w}(\rv - \bm{r}_{-})$ [this follows from coarse graining \refeq{eq:defineQ} with $Q_{\rv} = \delta_{\rv,\rv_+} - \delta_{\rv,\rv_-}$].

The general solution for the magnetic field is \cite{Maggs2002}
\begin{equation}
B_{\mu} = -\partial_{\mu}\phi + \epsilon_{\mu\nu}\partial_{\nu}h \punc ,
\end{equation}
where $\phi$ is fixed by the Poisson equation $\nabla^{2}\phi = -Q(\rv)$, and $h$ is a Coulomb phase height. After simplification, the action reads
\beq
S\sub{SDM} = \int d^{2}\bm{r} \, \frac{\kappa}{2} \left(\lvert \del\phi \rvert ^2 + \lvert \del h \rvert ^{2}\right)
\eeq
(cross terms vanish after integration by parts) and, since the second contribution returns $Z$, \refeq{eq:gmaction} reduces to
\begin{equation}
G\sub{m}(\bm{R}) = e^{-\int d^{2}\bm{r} \, \frac{\kappa}{2} \lvert \del\phi \rvert ^2} \punc .
\end{equation}

The remaining integral is the energy associated with an electrostatic potential $\phi$ due to two extended charge distributions $\pm\mathcal{K}_{w}$ separated by $\bm{R} = \rv_{+} - \rv_{-}$ \cite{Griffiths2017}. 
% actually, this is not quite true in two dimensions because asymptotic behavior of potentials is log(r) rather than 1/r and surface integral is not zero
For large monomer separation $\lvert \bm{R} \rvert \gg w$, the charge distributions `see' one another as point charges, hence (in two dimensions)
\begin{equation}
\int d^{2}\bm{r} \, \frac{1}{2}\lvert \bm{\nabla}\phi \rvert^{2} = \frac{1}{2\pi} \log \lvert \bm{R} \rvert \punc ,
\end{equation}
% two contributions: interaction of each distribution with other + interaction of each distribution with itself. The first is shown here, and the second is an additive (R independent) constant, which could in principle be calculated using the coarse-graining kernel
% also dropping infinities that appear in potential
up to additive constants, and the asymptotic behavior is
\begin{equation}
\label{eq:heightgm}
G\sub{m}(\bm{R}) \sim \lvert \bm{R} \rvert ^{-\kappa/2\pi} \punc .
\end{equation}

\subsection{DDM $G\sub{m}(\bm{R})$ in the Coulomb phase}
\label{ddmgminthecoulombphase}

By extension, the monomer distribution function in the DDM Coulomb phase, with a pair of monomers in one replica, say $\alpha = 1$, is given by
\begin{equation}
G\sub{m}(\bm{R}) = \frac{1}{Z}\int \mathcal{D}\bm{B}\spr+(\rv)\mathcal{D}\bm{B}\spr-(\rv) \, e^{-S\sub{DDM}[\bm{B}\spr+,\bm{B}\spr-]} \punc ,
\end{equation}
where $Z$ is the partition function in the close-packed case, $S\sub{DDM}$ is given by \refeq{eq:DDMCoulombB} and $\del \cdot \bm{B}\spr\pm = Q(\rv)$ [because $\bm{B}\spr\pm = \bm{B}\spr1 \pm \bm{B}\spr2$, $\del \cdot \bm{B}\spr1 = Q(\rv)$, and $\del \cdot \bm{B}\spr2 = 0$].

The calculation proceeds as in the previous section, now with two fields $\bm{B}\spr\pm$ and their stiffnesses $\kappa_{\pm}$, giving
\begin{equation}
G\sub{m}(\bm{R}) \sim \lvert \bm{R} \rvert ^{-\eta} \punc ,
\end{equation}
where $2\pi\eta = \kappa_{+} + \kappa_{-}$. As required, this reduces to \refeq{eq:heightgm} when $K=0$ [since then $\kappa_\pm = \frac{\kappa}{2}$ from \refeq{eq:kpm}].

\subsection{DDM $G\sub{d}(\bm{R})$ in the (anti)synchronized phases}
\label{ddmgdinthe(anti)synchronizedphases}

% Action
Finally, we calculate the double monomer distribution function in the DDM (anti)synchronized phases. In these phases, the cosine term in \refeq{eq:DDMsync} is relevant and locks the relative height to values $h\spr- = 0 \, (\frac{1}{2})$. Hence, from the continuum version of \refeq{eq:height}, the corresponding magnetic field $\bm{B}\spr- = 0$. For the total magnetic field, this implies $\bm{B}\spr+ = 2\bm{B}\spr1$, since the cosine term in \refeq{eq:DDMcol} is irrelevant. In this case, \refeq{eq:DDMCoulombB} reduces to
\begin{equation}
S\sub{DDM} = \int d^{2}\bm{r} \, 2\kappa_{+}\lvert \bm{B}\spr1 \rvert^2 \punc,
\end{equation}
which is the correct action for the (anti)synchronized phases. 

% Gd
In terms of this, the continuum version of the double monomer distribution function is
\begin{equation}
G\sub{d}(\bm{R}) = \frac{1}{Z}\int \mathcal{D}\bm{B}\spr1(\rv) \, e^{-S\sub{DDM}[\bm{B}\spr1]} \punc ,
\end{equation}
where $Z$ is the partition function in the close-packed case and $\bm{B}\spr1 = \bm{B}\spr2$ is the magnetic field in the presence of a pair of test monomers, i.e., $\del \cdot \bm{B}\spr1 = Q(\rv)$. The derivation proceeds as in \refapp{sdmgminthecoulombphase} but with $\kappa\rightarrow 4\kappa_{+}$, and the asymptotic behavior is
\begin{equation}
\label{eq:heightgd}
G\sub{d}(\bm{R}) \sim \lvert \bm{R} \rvert ^{-2\kappa_{+}/\pi} \punc .
\end{equation}

\bibliography{dimersbibliography}

\end{document}